\documentclass[12pt]{iopart}
\usepackage{mathrsfs}
\usepackage{graphicx,color}

\usepackage{iopams} 
\def\AS{\bar{V}}
\def\TM{{\bf T}} 
\def\HM{{\bf H}} 
\def\Aa{{\bf A}} 
\def\Ba{{\bf B}}
\def\Ca{{\bf C}}
\def\Da{{\bf D}}
\def\Cm{C}
\def\Cf{{\cal C}}
\def\Dm{D}
\def\Df{{\cal D}}
\def\Ff{{\cal F}}
\def\Qf{{\cal Q}}

\newcommand{\ket}[1]{\left| #1 \right\rangle}

\newcommand{\bra}[1]{\left\langle #1 \right|}

\def\lam{\lambda}

\def\lam{\lambda}

\def\ra{\rangle}
\def\la{\langle}
\def\ua{\uparrow}
\def\da{\downarrow}

\def\Tr{{\rm Tr\,}}
\def\bege{\begin{equation}}
\def\ende{\end{equation}} 
\begin{document}

\title[]{Derivation of Matrix Product Ansatz 
for the Heisenberg Chain from Algebraic Bethe Ansatz}

\author{Hosho Katsura${}^{1}$~and~
Isao Maruyama${}^{2}$}

\address{${}^1$ Kavli Institute for Theoretical Physics, University of California,
Santa Barbara, CA 93106, USA}
\address{${}^2$ Graduate School of Engineering Science, Osaka University, 
Toyonaka, Osaka 560-8531, Japan}
\ead{\mailto{katsura@kitp.ucsb.edu}, \mailto{maru@mp.es.osaka-u.ac.jp}}
\begin{abstract}
We derive a matrix product representation of the Bethe ansatz state for the XXX and 
XXZ spin-$\frac{1}{2}$ Heisenberg chains using the algebraic Bethe ansatz. 
In this representation, the components of the Bethe eigenstates are expressed as traces of products of matrices 
which act on ${\bar {\mathscr H}}$, the tensor product of auxiliary spaces. By changing the basis in ${\bar {\mathscr H}}$, 
we derive explicit finite-dimensional representations for the matrices. 
These matrices are the same as those appearing in the recently proposed
matrix product ansatz by Alcaraz and Lazo [Alcaraz F C and Lazo M J 2006 {\it J. Phys. A: Math. 
Gen.} \textbf{39} 11335.] apart from normalization factors. 
We also discuss the close relation between the matrix product representation of the Bethe eigenstates and 
the six-vertex model with domain wall boundary conditions [Korepin V E 1982
{\it Commun. Math. Phys.}, \textbf{86} 391.]
and show that the change of basis corresponds to a mapping from the six-vertex model to the five-vertex model. 

\end{abstract}

\submitto{\JPA}
\maketitle

\section{Introduction}
Matrix product states (MPS) have attracted considerable interest in the interdisciplinary field of
condensed matter physics and quantum information science~\cite{Verstraete-Cirac, Perez-Garcia}. 
The archetype of MPS can be traced back
to the seminal work of Affleck, Kennedy, Lieb, and Tasaki~\cite{AKLT_PRL, AKLT_CMP, Baxter}, which is aimed at understanding the ground
state properties of Haldane gap systems~\cite{Haldane_PLA, Haldane_PRL}. It was then formalized 
in more generalized and abstract ways~\cite{Fannes, Kluemper1, Kluemper2}. After those works, the density matrix renormalization group (DMRG)
method, a powerful numerical method to study one-dimensional strongly correlated systems, was introduced
by White~\cite{White1,White2} and its connection to MPS formulation was discussed by Rommer and \"Ostlund~\cite{Rommer}.
A common feature of mathematically rigorous and numerical approaches is that both have failed to
describe quantum critical ground states which exhibit quasi-long range order. In the context of quantum
information theory, Vidal {\it et al}., have attempted to characterize the quantitative difference between
non-critical and critical ground states in one dimension in terms of entanglement entropy~\cite{Vidal}. 
However, a natural question to ask is whether or not there are quantum critical ground states expressed in a form of MPS. 
Surprisingly, the answer is {\it yes}. In Ref.~\cite{Alcaraz1} and the subsequent work \cite{Alcaraz2}, Alcaraz and Lazo have actually shown that the eigenstates of the spin-$\frac{1}{2}$ Heisenberg chain
can be expressed as MPS. It has been well known that this model is gapless and the quasiparticles, 
so-called {\it spinons}, have a linear dispersion relation since the pioneering work of Bethe in 1931~\cite{Bethe}.
The method to solve this model used by Bethe is called coordinate Bethe ansatz and there are
several variants (algebraic, functional, etc.,) of it. Alcaraz and Lazo proposed another alternative
formulation of the Bethe ansatz in terms of MPS, which they call matrix product ansatz (MPA).
In this formulation, one can obtain the spectrum conditions, i.e., Bethe equations, imposing algebraic
relations between matrices consisting of MPS. The physical meaning of those matrices is
interpreted as the generators of the Zamolodchikov-Faddeev algebra in (1+1) dimensional field theories
~\cite{Zam}. In Ref.~\cite{Alcaraz-Lazo3}, Alcaraz and Lazo applied MPA to other integrable models such as the Hubbard model~\cite{Lieb-Wu}, fermionic supersymmetric $t$-$J$ model~\cite{Schlottmann}, and Fateev-Zamolodchikov model~\cite{Fateev-Zamolodchikov}, and obtained the correct Bethe equations. 

In this paper, we show that MPA is essentially equivalent to the algebraic Bethe ansatz (ABA) in the XXX and XXZ spin-$\frac{1}{2}$ Heisenberg chains. 
The ABA is an elegant method for solving the eigenvalue problem of quantum integrable models 
developed in the late 70s~\cite{Faddeev, QISM, Nepomechie}. 
So far, the relation between the MPS and ABA has been discussed in a completely different context, i.e.,
stochastic Markovian models in one dimension~\cite{Golinelli-Mallick}. 
One of the simplest examples of the models is the asymmetric simple exclusion process (ASEP), 
which plays an important role in non-equilibrium statistical mechanics. This model was first exactly 
solved {\it not} using the Bethe ansatz~\cite{Derrida} while the relation to the integrable 
quantum spin chains was then clarified by Alcaraz \etal., \cite{AlcarazDroz}. 
It was first discovered in Ref.~\cite{Golinelli-Mallick} that the MPA for ASEP can be derived from the ABA. 
The authors have also constructed the explicit finite dimensional matrices for the MPA. 
The key ingredient is the change of basis in ${\bar \mathscr{H}}$, the tensor product of auxiliary spaces. 
Since the MPS are defined as traces over ${\bar \mathscr{H}}$, they are invariant under the change  
and hence one can take an appropriate basis in which the matrices have very simple expressions. 
Along the same lines as their approach, we derive the MPA for the XXX and XXZ spin-$\frac{1}{2}$ Heisenberg chains. 
The explicit expressions for the matrices are also obtained. 

The organization of this work is as follows. In Section 2, we review the ABA method for solving the eigenvalue problem of the Heisenberg chain. In section 3, we derive the MPS representations for the Bethe eigenstates 
from the ABA by preparing ${\bar \mathscr{H}}$, the tensor product of the auxiliary spaces. 
In section 4, the explicit expressions for the matrices appearing in the MPA are obtained by solving the recursion relations for the matrices. In section 5, the relation between the MPA and the six-vertex model with domain wall boundary conditions~\cite{Korepin_DWBC} is discussed. 
Conclusions and future perspectives are given in the last section. 
In Appendices, we provide graphical representations for the matrices in the main text and discuss the mapping from the six-vertex model to a five-vertex model.  

\section{Algebraic Bethe ansatz for the XXZ spin-$\frac{1}{2}$ Heisenberg chain}
The XXZ spin-$\frac{1}{2}$ Heisenberg model with the periodic boundary condition is described by the following Hamiltonian: 
\begin{equation}
\HM_{\rm XXZ}=\sum^L_{i=1} 
\left\{  
		\sigma^x_i \sigma^x_{i+1} +\sigma^y_i \sigma^y_{i+1} +\Delta (\sigma^z_i \sigma^z_{i+1}-1)
\right\}, 
\label{eq:XXZ_ham}
\end{equation}
where $L$ denotes the total number of sites and $\sigma^\alpha_i$ ($\alpha=x,y,z$) are the Pauli matrices defined on the $i$th site. 
Here $\Delta$ is the anisotropy parameter and 
the particular cases $\Delta=1$ and $\Delta=0$ correspond to the XXX and XY chains, respectively. 


The eigenstates of this model can be constructed using the ABA~\cite{Faddeev, QISM, Nepomechie, Deguchi}. We shall briefly review this construction. 
In the ABA, the central object is the quantum $R$-matrix which is the solution of the Yang-Baxter equation. 
For the XXZ model, the $R$-matrix acting on $\mathbb{C}^2 \otimes \mathbb{C}^2$ is given by
\begin{equation}
R(\lambda)= \frac{1}{\sinh(\lambda+\eta)} \left( \begin{array}{cccc}
\sinh(\lambda+\eta) & 0 & 0 & 0 \\
0 & \sinh \lambda & \sinh \eta & 0 \\
0 & \sinh \eta & \sinh \lambda & 0 \\
0 & 0 & 0 & \sinh (\lambda+\eta) 
\end{array}\right),
\label{eq:R-op}
\end{equation}
where $\lambda$ is the spectral parameter and the relation between $\eta$ and the anisotropy $\Delta$ is given by $\Delta = \cosh \eta$. 
Next, we introduce the quantum $L$-operator represented by a matrix acting on the tensor product of two-dimensional vector spaces 
$\AS_j \otimes V_i$. 
The auxiliary space $\AS_j$ introduced here is spanned by two orthonormal states labelled as $\ket{\leftarrow}$ and $\ket{\rightarrow}$ while the physical Hilbert space at the $i$th site $V_i$ is spanned by $\ket{\uparrow}$ and $\ket{\downarrow}$.  
For the XXZ model, the $L$-operator ${\cal L}_{ji}(\lam)$ is defined by the $R$-matrix as 
\begin{equation}
{\cal L}_{ji}(\lam)=R_{ji} \left( \lam-\frac{\eta}{2} \right),
\end{equation} 
where ${\cal L}_{ji}$ acts on $\AS_j \otimes V_i$. 
Note that the operator ${\cal L}_{ji} (\lam)$ acts trivially on all the sites other than $V_i$. 
More explicitly, the $L$-operator is written as 
\begin{equation}
{\cal L}_{ji}(\lambda)=
\left(\begin{array}{cccc}
1 & 0 & 0 & 0\\
0 & b(\lambda) & c(\lambda) & 0\\
0 & c(\lambda) & b(\lambda) & 0\\
0 & 0 & 0 & 1
\end{array}\right),
\label{eq:L-op}
\end{equation}
where
\begin{equation}
b(\lambda)=\frac{\sinh\left(\lambda-\frac{\eta}{2}\right)}{\sinh\left(\lambda+\frac{\eta}{2}\right)},~~~~~c(\lambda)=\frac{\sinh \eta}{\sinh\left(\lambda+\frac{\eta}{2}\right)}
\end{equation}
for the XXZ chain and
\begin{equation}
b(\lambda)=\frac{\lambda-\frac{i}{2}}{\lambda+\frac{i}{2}}, ~~~~~ c(\lambda)=\frac{i}{\lambda+\frac{i}{2}}
\end{equation}
for the XXX chain. The latter can be obtained by taking the scaling limit of the former, i.e.,  $\lambda_{\rm XXZ}=\epsilon \lambda_{\rm XXX}$ and $\eta=i \epsilon$ with the limit of $\epsilon \to 0$. 
It is useful to note the relation between $b(\lambda)$, $c(\lambda)$ and the physical quasi-momentum $k$. For both the XXZ and XXX cases, $b(\lambda)=z$ with $z=e^{-ik}$~\cite{Alcaraz1}. By a straightforward calculation, one can verify the following relation:
\begin{equation}
c(\lambda)^2=1-2\Delta z +z^2. 
\label{important_relation}
\end{equation} 
In particular, $c(\lambda)=1-z$ when $\Delta=1$ corresponding to the XXX chain. 
The following identity can then be shown as a direct consequence of the Yang-Baxter equation:
\bege
R_{01}(\lambda-\mu) {\cal L}_{0i}(\lambda) {\cal L}_{1i}(\mu)={\cal L}_{1i}(\mu) {\cal L}_{0i}(\lambda) R_{01}(\lambda-\mu). 
\label{eq:RLL=LLR}
\ende
Here, $R_{01}$ acts nontrivially on $\AS_0 \otimes \AS_{1}$ and trivially on $V_i$. 
Note that ${\cal L}_{1i}$ acts nontrivially on $\AS_{1} \otimes V_i$ and trivially on $\AS_0$ and the other spatial sites. 

The monodromy matrix is then constructed as the following ordered matrix product: 
\bege
{\cal T}_0(\lam)={\cal L}_{01}(\lam) {\cal L}_{02} (\lam) ... {\cal L}_{0L}(\lam). 
\ende
In the basis of $\AS_0$, the monodromy matrix can be represented as a $2 \times 2$ matrix:
\begin{equation}
{\cal T}_0(\lam)=\left(\begin{array}{cc}
\Aa (\lam) & \Ba (\lam) \\
\Ca (\lam) & \Da (\lam) 
\end{array}\right), 
\end{equation}
where matrix elements $\Aa (\lam)$, $\Ba (\lam)$, $\Ca (\lam)$, and $\Da (\lam)$ are themselves operators acting on the total Hilbert space $\mathscr{H}=\otimes^L_{i=1} V_i$. 
Using the relation Eq. (\ref{eq:RLL=LLR}), one can show the following relation for the monodromy matrix. 
\bege
R_{01} (\lambda - \mu) {\cal T}_0 (\lambda) {\cal T}_{1} (\mu) ={\cal T}_{1} (\mu) {\cal T}_0 (\lambda) R_{01}(\lambda-\mu).
\label{eq:RTT=TTR}
\ende
Graphical representations for both Eq. (\ref{eq:RLL=LLR}) and (\ref{eq:RTT=TTR}) are shown in Appendix B.2.
The commutation relations among $\Aa$, $\Ba$, $\Ca$, and $\Da$ can be obtained from this relation. 
Taking the trace of the monodromy matrix over $\AS_0$, one obtains a one-parameter family of transfer matrices acting on ${\mathscr H}$: 
\begin{equation}
{\bf T}(\lambda)=\Tr_{\AS_0} {\cal T}_0(\lam) =\Aa (\lam) + \Da (\lam).
\end{equation}
The Hamiltonian Eq. (\ref{eq:XXZ_ham}) can then be obtained from ${\bf T}(\lambda)$ by the trace identity:
\bege
{\bf H}_{\rm XXZ}=2 \sinh \eta\, \frac{\partial}{\partial \lambda} \log {\bf T}(\lambda) \Big|_{\lambda=\eta/2} +{\rm const.}
\ende
Since the Hamiltonian commutes with the monodromy matrix, one can construct a simultaneous eigenstate of both $\bf{H}_{\rm XXZ}$ and $\bf {T}$. 
The eigenstate of this operator is constructed by $\Ba (\lam)$'s as 
\begin{equation}
|\lam_1, \lam_2...\lam_n\rangle = \Ba (\lam_n) ... \Ba (\lam_1) \Phi_0, 
\end{equation}
where $\Phi_0$ denotes the reference ferromagnetic state, i.e., $\Phi_0=\ket{\Uparrow} \equiv | \uparrow, \uparrow ... \uparrow \rangle$ and 
$n$ denotes the number of down spins. 
We call the above state a Bethe state. Since one can show that $\Ba (\lam_i)$'s commute each other from the relation (\ref{eq:RTT=TTR}), 
this state is invariant under permutations of $\lam_i$'s. 
The spectrum conditions, i.e., the Bethe equations, are then obtained using the commutation relations between $\Aa$, $\Da$, and $\Ba$~\cite{Faddeev, QISM, Nepomechie}. 
For the XXZ model, those equations are given by
\bege
\left( \frac{\sinh(\lambda_j-\frac{\eta}{2})}{\sinh(\lambda_j+\frac{\eta}{2})} \right)^L \prod^n_{k=1 \atop k \ne j} \frac{\sinh(\lambda_j-\lambda_k+\eta)}{\sinh(\lambda_j-\lambda_k-\eta)}=1,~~~j=1, 2, ..., n. 
\ende


\section{Derivation of the matrix product state representation from the algebraic Bethe ansatz}
In the previous section, we outlined the construction of the eigenstates of $\bf{H}_{\rm XXZ}$ using the ABA. 
In this section, we derive the matrix product state representations for the eigenstates from the ABA. 
Let $\ket{\leftarrow}$ and $\ket{\rightarrow}$ be two orthonormal states spanning $\AS$. 
Then the Bethe state is expressed as
\begin{eqnarray}
|\lam_1, \lam_2...\lam_n\rangle 
                            &=& \Ba(\lam_n) ... \Ba(\lam_2) \Ba(\lam_1) \Phi_0 \nonumber \\
                            &=& \bra{\leftarrow}{\cal T}(\lam_n)\ket{\rightarrow} ... \bra{\leftarrow}{\cal T}(\lam_2)\ket{\rightarrow} \bra{\leftarrow}{\cal T}(\lam_1)\ket{\rightarrow} \Phi_0 \nonumber \\
                            &=& \Tr_{{\AS}^{\otimes n}} (Q_n {\cal T}(\lam_n)\otimes ...\otimes {\cal T}(\lam_2) \otimes {\cal T}(\lam_1) \Phi_0) \nonumber \\
                            &=& \Tr_{{\AS}^{\otimes n}} \left( Q_n \left[ \prod^{L}_{i=1} {\cal L}_i (\lam_1, ..., \lam_n)\right] \Phi_0 \right)
\label{eq:BetheStateMPA}
\end{eqnarray}
with $Q_n=\ket{\rightarrow, \rightarrow ... \rightarrow} \bra{\leftarrow, \leftarrow ... \leftarrow} \equiv |\Rightarrow\ra \la \Leftarrow|$ and ${\cal L}_i (\lam_1, ..., \lam_n) \equiv {\cal L}_i (\lam_n) \otimes \cdots \otimes {\cal L}_i (\lam_1)$ which acts on the vector space $V_i$ only in $\mathscr{H}$. 
Here, we have omitted the indices for the auxiliary spaces. 
It would be helpful to note that the following identity for tensor products holds: 
$(A.B)\otimes(C.D)=(A\otimes C). (B \otimes D)$. 

We now introduce two matrices $D_n$ and $C_n$ via 
\begin{equation}
{\cal L}_i (\lam_1, ..., \lam_n)|\ua \ra =D_n (\lam_1, ..., \lam_n)|\ua\ra + C_n(\lam_1, ..., \lam_n)|\da\ra 
\end{equation}
or equivalently, $D_n(\lam_1, ..., \lam_n)=\la \ua| {\cal L}_i (\lam_1, ..., \lam_n)|\ua \ra$ and 
$C_n(\lam_1, ..., \lam_n)=\la \da | {\cal L}_i (\lam_1, ..., \lam_n) |\ua\ra$. 
It should be noted that $D_n, C_n$, and $Q_n$ are $2^n \times 2^n$ matrices acting on ${\AS}^{\otimes n}$ with scalar elements. 
The first terms are, for example, given by
\begin{equation}
D_1(\lam)=\left(\begin{array}{cc}
1 & 0 \\
0 & b(\lam) 
\end{array}\right),~~
C_1(\lam)=\left(\begin{array}{cc}
0 & c(\lam) \\
0 & 0 
\end{array}\right),~~{\rm and}~~
Q_1(\lam)=\left(\begin{array}{cc}
0 & 0 \\
1 & 0 
\end{array}\right).
\label{eq:first terms}
\end{equation}
Henceforth, we shall denote the $n$-fold tensor product of the auxiliary spaces as ${\bar {\mathscr H}} \equiv \bigotimes^n_{j=1} {\bar V}_{n-j+1}$. 
Using $D_n, C_n$, and $Q_n$, the Bethe state can be written as
\begin{equation}
|\lam_1, \lam_2, ..., \lam_n\ra = \Tr_{\bar {\mathscr H}} \left[ Q_n \prod^L_{i=1} (D_n(\lam_1, ..., \lam_n)|\ua \ra_i+ C_n(\lam_1, ..., \lam_n)|\da \ra_i) \right], 
\label{eq:MPS form of Bethe state}
\end{equation}
where $|\ua\ra_i$ and $|\da \ra_i$ denote the up and down spin states at the $i$th site, respectively. 
This form can be regarded as the matrix product state in the usual sense except for the boundary matrix $Q_n$. 

Let us consider the recursion relation between $D_{n+1}$, $C_{n+1}$ and $D_n$, $C_n$. 
From the fact ${\cal L}_i (\lam_1, ..., \lam_n, \lam_{n+1})={\cal L}_i(\lam_{n+1})\otimes {\cal L}_i (\lam_1, ..., \lam_n)$, 
one can derive the following relations:
\begin{eqnarray}
\fl 
D_{n+1}(\lam_1 ...\lam_n, \lam_{n+1}) &= \left(\begin{array}{cc} 1 & 0 \\ 0 & b(\lam_{n+1}) \end{array} \right) \otimes D_n (\lam_1 ... \lam_n)
+ \left(\begin{array}{cc} 0 & 0\\ c(\lam_{n+1}) & 0 \end{array}\right) \otimes C_n(\lam_1 ... \lam_n) \nonumber \\
                                     &= \left(\begin{array}{cc} D_n(\lam_1 ... \lam_n) & 0 \\ c(\lam_{n+1})C_n(\lam_1 ... \lam_n) & b(\lam_{n+1}) D_n(\lam_1 ... \lam_n) \end{array} \right) 
\label{recD}
\end{eqnarray}
and
\begin{eqnarray}
\fl 
C_{n+1}(\lam_1 ... \lam_n, \lam_{n+1}) &= \left(\begin{array}{cc} 0 & c(\lam_{n+1}) \\ 0 & 0 \end{array}\right) \otimes D_n(\lam_1 ... \lam_n)
+ \left(\begin{array}{cc} b(\lam_{n+1}) & 0\\ 0 & 1 \end{array}\right) \otimes C_n(\lam_1 ... \lam_n) \nonumber \\
&= \left(\begin{array}{cc} b(\lam_{n+1}) C_n(\lam_1 ... \lam_n) & c(\lam_{n+1}) D_n(\lam_1 ... \lam_n) \\ 0 & C_n(\lam_1 ... \lam_n)  \end{array}\right). 
\label{recE}
\end{eqnarray}
Here we have used the following identities: 
\begin{eqnarray}
\fl~~~~
{}_i \la \ua| {\cal L}_i(\lam_{n+1}) |\ua \ra_i = \left( \begin{array}{cc}
1 & 0 \\ 0 & b(\lam_{n+1})
\end{array} \right),~~~
{}_i \la \ua| {\cal L}_i(\lam_{n+1}) |\da \ra_i = \left( \begin{array}{cc}
0 & 0 \\  c(\lam_{n+1}) & 0
\end{array} \right), \nonumber \\
\fl~~~~
{}_i \la \da| {\cal L}_i(\lam_{n+1}) |\ua \ra_i = \left( \begin{array}{cc}
0 & c(\lam_{n+1}) \\ 0 & 0
\end{array} \right),~~~
{}_i \la \da| {\cal L}_i(\lam_{n+1}) |\da \ra_i = \left( \begin{array}{cc}
b(\lam_{n+1}) & 0 \\  0 & 1
\end{array} \right). \nonumber 
\end{eqnarray}
The above matrices are understood to act on the auxiliary space ${\bar V}_i$ 
(see Appendix B.1 for a more detailed explanation).  
By definition, it is obvious that $Q_n(\lam_1... \lam_n, \lam_{n+1} )$ satisfies the following recursion relation:
\begin{equation}
Q_{n+1}(\lam_1 ... \lam_n, \lam_{n+1} )
=\left(\begin{array}{cc} 0 & 0 \\ Q_n(\lam_1 ... \lam_n) & 0 \end{array}\right) \label{recQ}
\end{equation}
and hence the explicit form of $Q_n (\lam_1, ... ,\lam_n)$ is given by
\begin{equation}
Q_n(\lam_1, ..., \lam_n) = \bigotimes^n_{l=1} 
\left(\begin{array}{cc} 0 & 0 \\ 1 & 0 \end{array}\right). 
\end{equation}
Therefore, one can construct a matrix product representation of the Bethe ansatz state using the relations (\ref{recD}-\ref{recQ}) recursively. 
The interesting point here is that the dimension of the matrix is finite even in the thermodynamic limit 
if the number of down spins is finite. On the other hand, if we consider the case of the fixed magnetization, 
i.e., the ratio $n/L$ is fixed, the dimension of the matrix becomes infinity in the limit of $L \to \infty$. 


\section{Change of the basis in $\bar{\mathscr{H}}$}
In the previous section, we have derived the matrix product state representation for the Bethe states from the ABA.
At that point, however, the matrices are defined by the recursion relations and the connection to another representation 
proposed by Alcaraz and Lazo~\cite{Alcaraz1, Alcaraz2} is not clear. In this section, we clarify the direct relation between them by solving the recursion relations. 
We now try to rewrite the matrices $D_n$ and $C_n$ as diagonally as possible. 
For simplicity, we omit here the indeterminants of matrices, i.e., $\lam_1, ..., \lam_n$. 
Furthermore, we will replace $b(\lambda_j)$ with $z_j=e^{-ik_j}$. 
Suppose that $D_n$ is diagonalized by the invertible matrix $F_n$ as $F^{-1}_n D_n F_n = {\Df}_n$. 
In the new basis in ${\bar {\mathscr H}}$, $C_n$ and $Q_n$ are transformed into ${\Cf}_n=F^{-1}_n C_n F_n$ and ${\cal Q}_n=F^{-1}_n Q_n F_n$, respectively. 
From the cyclic property of the trace, it is obvious that the Bethe state can be written by new matrices as
\bege
|\lam_1, \lam_2, ..., \lam_n\rangle = \Tr_{\bar {\mathscr H}} \left[ {\cal Q}_n \prod^L_{i=1} (\Df_n |\uparrow\rangle_i+\Cf_n |\downarrow \rangle_i) \right]. 
\label{MPS_in_new_basis}
\ende 
We now decompose ${\Cf}_n$ as a sum of $n$ matrices, 
\bege
{\Cf}_n=\sum^n_{i=1}{\Cf}^{(i)}_n, 
\label{decomposition}
\ende
and suppose that $\Df_n$ and ${\Cf}^{(i)}_n$ satisfy the following algebraic relations: 
\begin{eqnarray}
{\Cf}^{(i)}_n \Df_n = z_i \Df_n {\Cf}^{(i)}_n \label{alg1}\\
{\Cf}^{(i)}_n {\Cf}^{(j)}_n = {\tilde S}_{ij}(z_i, z_j) {\Cf}^{(j)}_n {\Cf}^{(i)}_n \label{alg2}\\
{\Cf}^{(i)}_n {\Cf}^{(i)}_n = 0. \label{alg3} 
\end{eqnarray}
Then we look for an appropriate ${\tilde S}_{ij}(z_i, z_j)$ by observation. 
Let us now consider the case of $n=2$. In this case, $D_2$ and $C_2$ are given by
\begin{eqnarray}
D_2=
\left(\begin{array}{cc} D_1(\lam_1) & 0 \\ 0 & z_2 D_1(\lam_1) \end{array}\right) 
+\left(\begin{array}{cc} 0 & 0 \\ c(\lam_2) C_1(\lam_1) & 0 \end{array}\right), 
\\
C_2= 
\left(\begin{array}{cc} 0 & c(\lam_2) D_1(\lam_1) \\ 0 & 0 \end{array}\right)
+\left(\begin{array}{cc} z_2 C_1(\lam_1) & 0 \\ 0 & C_1(\lam_1) \end{array}\right),
\end{eqnarray}
respectively, where $D_1(\lam_1)$ and $C_1(\lam_1)$ are defined in Eq. (\ref{eq:first terms}). By an explicit calculation, one can confirm that the matrix $D_2$ is diagonalized as
\begin{eqnarray}
{\cal D}_2 = F^{-1}_2 D_2 F_2 = \left( \begin{array}{cc} D_1(\lam_1) & 0 \\ 0 & z_2 D_1(\lam_1) \end{array}\right), 
\end{eqnarray}
where 
\bege
F_2 = \left(\begin{array}{cc} 1 & 0 \\ {\cal F}_1 & 1 \end{array}\right)
\ende
with ${\cal F}_1 = \frac{c(\lam_2)}{z_1-z_2} C_1(\lam_1)$. We note that the inverse of $F_2$ is given by 
\bege
F^{-1}_2 = \left(\begin{array}{cc} 1 & 0 \\ -{\cal F}_1 & 1 \end{array}\right). 
\ende
In the new basis defined by $F_2$, the matrix $C_2$ becomes 
\bege
{\cal C}_2 = F^{-1}_2 C_2 F_2 =
\left( \begin{array}{cc} z_2 C_1(\lam_1) + c(\lam_2) {\cal F}_1 & c(\lam_2) D_1(\lam_1) \\0 & -z_1 c(\lam_2) {\cal F}_1 +C_1(\lam_1) \end{array} \right). 
\ende
We now divide the above matrix into two matrices as ${\cal C}_2={\cal C}^{(1)}_2+{\cal C}^{(2)}_2$ with
\bege
\fl 
{\cal C}^{(1)}_2=\left(\begin{array}{cc} z_2 C_1(\lam_1)+c(\lam_2) {\cal F}_1 & 0 \\ 0 & -c(\lam_2) z_1 {\cal F}_1 +C_1(\lam_1) \end{array}\right), 
{\cal C}^{(2)}_2 = \left( \begin{array}{cc} 0 & c(\lam_2) D_1(\lam_1) \\ 0 & 0 \end{array}\right). 
\ende
One can confirm that ${\cal C}^{(1)}_2$ and ${\cal C}^{(2)}_2$ satisfy the algebraic relations (\ref{alg1}) and (\ref{alg3}). 
Then we compare the commutation relation between ${\cal C}^{(1)}_2$ and ${\cal C}^{(2)}_2$ with Eq. (\ref{alg2}) and find that 
${\tilde S}_{12}(z_1,z_2)$ should be given by
\bege
{\tilde S}_{12}(z_1,z_2)=-\frac{z_1}{z_2} \cdot \frac{z_1 z_2+1-2\Delta z_2}{z_1 z_2+1-2\Delta z_1}.  
\ende

From this observation, we take the coefficient ${\tilde S}_{ij}(z_i, z_j)$ to be 
\bege
{\tilde S}_{ij}(z_i,z_j)=-\frac{z_i}{z_j} \cdot \frac{z_i z_j+1-2\Delta z_j}{z_i z_j+1-2\Delta z_i}  
\label{eq:Stilde}
\ende
for general $i$ and $j$. 
We shall prove the algebraic relations (\ref{alg1}-\ref{alg3}) with this ${\tilde S}_{ij}(z_i, z_j)$ by induction on $n$. 
For $n=1$, the matrices ${\cal D}_1=D_1$, ${\cal C}_1=C_1$ trivially satisfy the relations (\ref{alg1}-\ref{alg3}). 
Now, we suppose that we have already found the matrix $F_n$ which diagonalizes $D_n$ as ${\cal D}_n=F^{-1}_n D_n F_n$ 
and have found a decomposition ${\cal C}_n=\sum^n_{i=1} {\cal C}^{(i)}_n$ such that the relations (\ref{alg1}-\ref{alg3}) are satisfied 
between ${\cal D}_n$ and ${\cal C}^{(i)}_n$'s. Then we show that it is possible to construct a matrix $F_{n+1}$ which diagonalizes $D_{n+1}$ 
and can find a decomposition of ${\cal C}_{n+1}$. 
First, we take the matrix $F_{n+1}$ to be of the following form:
\begin{equation}
F_{n+1} = \left(\begin{array}{cc} F_n & 0 \\ F_n {\cal F}_n & F_n \end{array}\right)
\end{equation}
or equivalently 
\begin{equation}
F^{-1}_{n+1} = \left(\begin{array}{cc} F^{-1}_n & 0 \\ -{\cal F}_n F^{-1}_n & F
^{-1}_n \end{array}\right).  
\end{equation}
We then obtain the following recursion relations using Eqs. (\ref{recD}-\ref{recQ}):
\begin{eqnarray}
\fl 
\Df_{n+1}
= \left(\begin{array}{cc} \Df_n & 0 \\ -{\cal F}_n \Df_n +c(\lam_{n+1}){\Cf}_n +z_{n+1}\Df_n {\cal F}_n & z_{n+1}\Df_n
 \end{array}\right), \label{recD2}\\ 
 \fl 
{\Cf}_{n+1} 
= \left(\begin{array}{cc} z_{n+1}{\Cf}_n+c(\lam_{n+1}) \Df_n {\Ff}_n & c(\lam_{n+1})\Df_n \\  
{\Cf}_n {\Ff}_n -z_{n+1} {\Ff}_n  {\Cf}_n -c(\lam_{n+1}) {\Ff}_n \Df_n {\Ff}_n & {\Cf}_n-c(\lam_{n+1}){\cal F}_n 
\Df_n \end{array} \right) \label{recE2}, \\
\fl 
{\cal Q}_{n+1}
=\left(\begin{array}{cc} 0 & 0 \\ {\cal Q}_n & 0 \end{array}\right), \label{recQ2}
\end{eqnarray}
where ${\cal F}_n$ is an unknown matrix to be determined. 
Since one can take $F_1$ to be the $2 \times 2$ identity matrix, ${\cal Q}_1=Q_1$ holds and hence ${\cal Q}_n=Q_n$ for ${}^\forall n$. 
In other words, the change of basis preserves the domain wall boundary condition (DWBC) in ${\bar \mathscr{H}}$. This property plays a crucial role when we interpret our results in terms of the six-vertex model as discussed in Sec. 5. 
From Eq. (\ref{recD2}), one can see that $\Df_{n+1}$ is a diagonal matrix 
if the matrix $\Ff_n$ satisfies $-\Ff_n \Df_n +c(\lam_{n+1}){\Cf}_n +z_{n+1}\Df_n \Ff_n=0$.  
Using Eq. (\ref{alg1}), such $\Ff_n$ can be constructed as
\begin{equation}
\Ff_n =  c(\lam_{n+1}) \Df^{-1}_n \sum^n_{i=1} \frac{{\Cf}^{(i)}_n}{z_i-z_{n+1}}.  
\label{Ff_n}
\end{equation}
Next, we shall show that the (2,1)-component of ${\Cf}_{n+1}$ is also zero if ${\tilde S}_{ij}(z_i, z_j)$ is given by Eq. (\ref{eq:Stilde}). 
By a direct calculation, one obtains 
\begin{eqnarray}
&& {\Cf}_n \Ff_n -z_{n+1} \Ff_n  {\Cf}_n -c(\lam_{n+1}) \Ff_n \Df_n \Ff_n \nonumber \\
&=& c(\lam_{n+1}) \Df^{-1}_{n} \sum^n_{i=1} \sum^n_{j=1} d_{ij} \Cf^{(i)}_n \Cf^{(j)}_n
\label{showzero}
\end{eqnarray}
with 
\begin{eqnarray}
d_{ij} &=& z^{-1}_i \frac{1}{z_j-z_{n+1}} -z_{n+1} \frac{1}{z_i-z_{n+1}} -\frac{c(\lam_{n+1})^2}{(z_i-z_{n+1})(z_j-z_{n+1})} \nonumber \\
&=& z^{-1}_i \frac{1}{z_j-z_{n+1}} -z_{n+1} \frac{1}{z_i-z_{n+1}} -\frac{1-2 \Delta z_{n+1}+z^2_{n+1}}{(z_i-z_{n+1})(z_j-z_{n+1})} \nonumber \\
&=& -\frac{z_{n+1}}{z_i}\cdot \frac{z_i z_j+1-2 \Delta z_i}{(z_i-z_{n+1})(z_j-z_{n+1})}, 
\end{eqnarray}
where we have used the relation Eq. (\ref{important_relation}). 
Using the relations (\ref{alg2}) and (\ref{alg3}), Eq. (\ref{showzero}) is zero if $d_{ij}$ satisfies $d_{ij}+[{\tilde S}_{ij}(z_i, z_j)]^{-1} d_{ji}=0$. 
Although this relation is highly nontrivial, one can confirm that it holds if ${\tilde S}_{ij}(z_i, z_j)$ is given by Eq. (\ref{eq:Stilde}). 
The recursion relations Eqs. (\ref{recD2}) and (\ref{recE2}) can now be written as
\begin{eqnarray}
\Df_{n+1}
&=& \left(\begin{array}{cc} 1 & 0 \\ 0 & z_{n+1} \end{array}\right) \otimes \Df_n, \label{recD3}\\
{\Cf}_{n+1}
&=& \left(\begin{array}{cc} \sum^n_{i=1}\frac{z_i z_{n+1}+1-2\Delta z_{n+1}}{z_i-z_{n+1}}{\Cf}^{(i)}_n & c(\lam_{n+1})\Df_n \\ 0 & -z_{n+1} \sum^n_{i=1} \frac{z_i z_{n+1} +1 -2\Delta z_i}{z_i-z_{n+1}} {\Cf}^{(i)}_n \end{array}\right). \label{recE3}
\end{eqnarray}
From them we deduce the decomposition of ${\Cf}_{n+1}=\sum^{n+1}_{i=1} {\Cf}^{(i)}_{n+1}$ with
\bege
\fl 
{\Cf}^{(i)}_{n+1} = \frac{1}{z_i-z_{n+1}}\left(\begin{array}{cc} (z_i z_{n+1}+1-2\Delta z_{n+1}){\Cf}^{(i)}_n & 0 \\ 0 & -z_{n+1}(z_i z_{n+1} +1 -2\Delta z_i){\Cf}^{(i)}_n \end{array}\right) ~~{\rm for}~~ i\le n, \label{recC3_1}
\ende
and
\bege
{\Cf}^{(n+1)}_{n+1} = \left(\begin{array}{cc} 0 & c(\lam_{n+1})\Df_n \\ 0 & 0 \end{array}\right). \label{recC3_2}
\ende
From the above expressions, it is shown that $\Df_{n+1}$ and ${\Cf}^{(i)}_{(n+1)}$ $(1 \le i \le n+1)$ also satisfy 
the algebraic relations Eqs. (\ref{alg1})-(\ref{alg3}). Therefore, we have proved that there exists a decomposition of $\Cf_n$ 
with algebraic relations (\ref{alg1}-\ref{alg3}) at any $n$.

We now write down the explicit expressions for $\Df_n$ and $\Cf^{(i)}_n$ using the recursion relations (\ref{recD3}-\ref{recC3_2}), 
From Eq. (\ref{recD3}), we first obtain
\bege
\Df_n = \bigotimes^n_{l=1} \left(\begin{array}{cc} 1 & 0 \\ 0 & z_{n-l+1} \end{array}\right). 
\label{newDf}
\ende
Then, from Eq. (\ref{recC3_2}), we obtain
\bege
\Cf^{(n)}_n = c(\lam_n) \left(\begin{array}{cc} 0 & 1 \\ 0 & 0 \end{array}\right) \bigotimes^n_{l=2} \left(\begin{array}{cc} 1 & 0 \\ 0 & z_{n-l+1} \end{array}\right).
\label{newCf_1}
\ende
We finally find the explicit form of $\Cf^{(i)}_n$ with $1 \le i \le n$. The relation (\ref{recC3_1}) can be recast as 
\bege
\Cf^{(i)}_{n+1} = -\frac{z_i z_{n+1} +1-2 \Delta z_i}{z_i-z_{n+1}} \left(\begin{array}{cc} S_{i,n+1}(z_i, z_{n+1}) & 0\\ 0 & z_{n+1} \end{array}\right) \otimes \Cf^{(i)}_n, 
\ende
where
\bege
S_{ij}(z_i,z_j)=-\frac{z_i z_j +1 -2\Delta z_j}{z_i z_j +1-2\Delta z_i}. 
\ende
Therefore, we obtain the explicit expression for $\Cf^{(i)}_n$ as
\bege
\fl 
\Cf^{(i)}_n = f(z_i,...,z_{n}) c(\lam_i) 
\bigotimes^{n-i}_{l=1} \left(\begin{array}{cc} S_{i,n-l+1}(z_i, z_{n-l+1}) & 0 \\ 0 & z_{n-l+1} \end{array}\right) \otimes 
\left(\begin{array}{cc} 0 & 1 \\ 0 & 0 \end{array}\right) \bigotimes^n_{l=n-i+2} \left(\begin{array}{cc} 1 & 0 \\ 0 & z_{n-l+1} \end{array}\right).
\label{newCf_2}
\ende
with
\bege
f(z_i,..., z_n)= \prod^n_{l=i+1} \left(-\frac{z_i z_l +1-2 \Delta z_i}{z_i-z_l}\right)
\ende
for $i=1, ..., n-1$. We note $f(z_n)=1$. 
The action of the matrix $\Cf^{(i)}_n$ is almost diagonal in ${\bar {\mathscr H}}$ except in the single auxiliary space $\AS_i$. 
Therefore, it is now quite trivial to confirm the algebraic relations (\ref{alg1}-\ref{alg3}) using Eqs. (\ref{newDf}), (\ref{newCf_1}) and (\ref{newCf_2}). 

Let us now clarify the relations between the ABA and the MPA explicitly. 
Keeping the notations as close as possible to those in~\cite{Alcaraz1, Alcaraz2}, 
the Bethe eigenstate in the MPA is written in the form:
\bege
\fl 
|\psi_{n}\rangle = \sum_{\{ x_1, x_2, ..., x_n \}} \Tr(E^{x_1-1}_n A_n E^{x_2-x_1-1}_n A_n \cdots E^{x_n-x_{n-1}-1}_n A_n E^{L-x_n}_n \Omega_n) |x_1, x_2, ..., x_n \rangle,   
\label{eq:MPA}
\ende
where $|x_1, x_2, ..., x_n\rangle$ ($1 \le x_1 < x_2 < ... < x_n \le L$) denote the configurations with down spins at $(x_1, x_2, ..., x_n)$ and the subscript $n$ indicates the total number of down spins. 
Note that in the original papers by Alcaraz and Lazo~\cite{Alcaraz1, Alcaraz2}, $(x_1, x_2, ..., x_n)$ are the locations of up spins. 
The matrix $A_n$ is decomposed by $n$ matrices as
\bege
A_n=\sum^{n}_{i=1} A_{k_i,n} E_n,
\ende
where the matrices $A_{k_i,n}$ obey the commutation relations
\begin{eqnarray}
A_{k_i,n} E_n = z_i E_n A_{k_i,n}, \label{AL_alg1}\\
A_{k_i,n} A_{k_j,n} = S_{ij}(z_i,z_j) A_{k_j,n} A_{k_i,n}, \label{AL_alg2} \\
E_n \Omega_n = e^{-iP} \Omega_n E_n, \label{AL_alg3}
\end{eqnarray}
with $P=\sum^n_{i=1} k_i$. 
The above relations assure that $|\psi_n\rangle$ is the eigenstate of the Heisenberg Hamiltonian  $\HM_{\rm XXZ}$. 
From Eq. (\ref{AL_alg2}), one can derive $A_{k_i,n}^2=0$. 
The relations (\ref{AL_alg1}-\ref{AL_alg2}) together with $A_{k_i,n}^2=0$ look very similar to Eqs. (\ref{alg1}-\ref{alg3}). 
We now try to find a one-to-one correspondence between the matrices in Eqs. (\ref{alg1}-\ref{alg3}) and those in Eqs. (\ref{AL_alg1}-\ref{AL_alg2}). To reproduce the correct commutation relations, the following relations are required:
\begin{equation}
E_n=\alpha \Df_n, ~~ \Omega_n=\beta {\cal Q}_n=\beta Q_n,~~{\rm and}~~ A_{k_i,n} E_n = \gamma_i {\Cf}^{(i)}_n, 
\label{correspondence MPA ABA}
\end{equation}
where $\alpha$, $\beta$, and $\gamma_i$ ($i=1, ..., n$) can be the arbitrary numbers. 
Here we have used the fact that $z_j {\tilde S}_{ij}(z_i,z_j)= z_i S_{ij}(z_i,z_j)$. 
For simplicity, let us fix $\alpha=\beta=1$. Then, 
from the above correspondence, we obtain $A_{k_i,n} =\gamma_i {\Cf}^{(i)}_n \Df^{-1}_n$ and find
\begin{eqnarray}
\fl 
E_n = \bigotimes^n_{l=1} \left(\begin{array}{cc} 1 & 0 \\ 0 & z_{n-l+1} \end{array}\right)
\\
\fl 
A_{k_i,n} =\gamma_i \frac{c(\lam_i)}{z_i} f(z_i, ..., z_n) \bigotimes^{n-i}_{l=1} \left(\begin{array}{cc} S_{i,n-l+1}(z_i, z_{n-l+1}) & 0 \\ 0 & 1 \end{array}\right) \otimes 
\left(\begin{array}{cc} 0 & 1 \\ 0 & 0 \end{array}\right) \bigotimes^n_{l=n-i+2} \left(\begin{array}{cc} 1 & 0 \\ 0 & 1 \end{array}\right), 
\label{A^n_k} \nonumber \\
\end{eqnarray}
which are indeed equivalent to the matrices found in~\cite{Alcaraz2} if we change the ordering of quasi-momenta from 
$(k_1, k_2, ..., k_n)$ to $(k_n, ..., k_2, k_1)$ and take an appropriate set of $\gamma_i$'s. 
In this way, we have derived the matrices appearing in the MPA using the ABA. Therefore, the matrix product Bethe ansatz is equivalent to the algebraic Bethe ansatz. 
We note here that if we take the XY limit ($\Delta=0$), $S(z_i, z_j)=-1$ and hence Eq. (\ref{A^n_k}) can be regarded as the Jordan-Wigner transformation in the {\it auxiliary} space ${\bar {\mathscr H}}$.

We remark on the coefficient $\gamma_i$ in the proportional relation between $A_{k_i,n} E_n$ and $\Cf^{(i)}_n$. 
In the original work by Alcaraz and Lazo, $A_{n}$ is defined by $\sum^n_{i=1} A_{k_i,n}E_n=\sum^n_{i=1} \gamma_i \Cf^{(i)}_n$, while ${\Cf}_n$ is defined by Eq. (\ref{decomposition}). 
Therefore, one may think that there is a constraint on the proportionality coefficients $\gamma_i$'s.   
However, they can be arbitrary.  Let us explain the reason for it. 
Since $A_{k_i,n}$'s are nilpotent ($A^2_{k_i,n}=0$), only the following products appear in Eq. (\ref{eq:MPA}):
\begin{equation}
\Omega_n E^{x_1-1}_n A_{k_{\sigma (1)},n} 
E^{x_2-x_1}_n A_{k_{\sigma (2)},n} \cdots E^{x_n-x_{n-1}}_n A_{k_{\sigma (n)},n} E^{L-x_n+1}_n
\label{AL_MP}
\end{equation}
where $\sigma$ are permutations of $(1,2,...,n)$.  
Therefore, if we take an arbitrary set of $\gamma_i$'s, 
the matrix product Eq. (\ref{AL_MP}) for any $\sigma$ has 
the same prefactor and hence the state is uniquely determined apart from an overall factor. 

We also remark on the relation between Eqs. (\ref{MPS_in_new_basis}) and (\ref{eq:MPA}). 
Using the correspondence (\ref{correspondence MPA ABA}), we can rewrite Eq. (\ref{MPS_in_new_basis}) as
\bege
|\lam_1, \lam_2, ..., \lam_n \ra= {\rm Tr}_{\bar{\mathscr H}} \left[ \Omega_n \prod^L_{i=1} (E_n|\ua\ra_i + A_n |\da\ra_i) \right],
\label{ABA to MPA}
\ende
where we have set $\alpha=\beta=\gamma_i=1$. 
It is easy to show $\ket{\lam_1, \lam_2, ..., \lam_n} = \ket{\psi_n}$, where $\ket{\psi_n}$ is defined in Eq. (\ref{eq:MPA}). 
One may think that in the expansion of the r.h.s. of Eq. (\ref{ABA to MPA}) 
there are traces of the following matrix products
\bege
\fl 
~~~~~~~~
\Omega_n E^{x_1-1}_n A_{k_{\sigma (i_1)},n} 
E^{x_2-x_1}_n A_{k_{\sigma (i_2)},n} \cdots E^{x_m-x_{m-1}}_n A_{k_{\sigma (i_m)},n} E^{L-x_m+1}_n,
~~~(m<n)
\label{eq:MPA_2}
\ende
where $(i_1, i_2, ..., i_m)$ are subsets of $(1,2,...,n)$. 
Such terms do not appear in Eq. (\ref{eq:MPA}). 
In fact, they are forbidden since ${\rm Tr}_{\bar {\mathscr H}} [\Omega_n M_n]$ is nonzero only when the matrix $M_n$ is written as
$M_n=\Omega^T_n+ \cdots$, where ${}^T$ denotes matrix transpose. Let us explain it in more detail. From Eq. (\ref{A^n_k}), it is obvious that only a single arrow which is one of the orthonormal vectors in ${\bar V}_i$ is flipped by the action of $A_{k_i,n}$. On the other hand,  ${\rm Tr}_{\bar {\mathscr H}} [\Omega_n M_n]$ can be recast as $\la \Leftarrow|M_n|\Rightarrow \ra$. This matrix element is nonzero only when all the arrows are flipped by the action of $M_n$. This proves that the trace of (\ref{eq:MPA_2}) is zero when $m < n$. 


\section{ Relation to the six-vertex model with domain wall boundary conditions}
\begin{figure}[htb]
\begin{center}
\vspace{.5cm}
\hspace{-.0cm}\includegraphics[width=0.9\columnwidth]{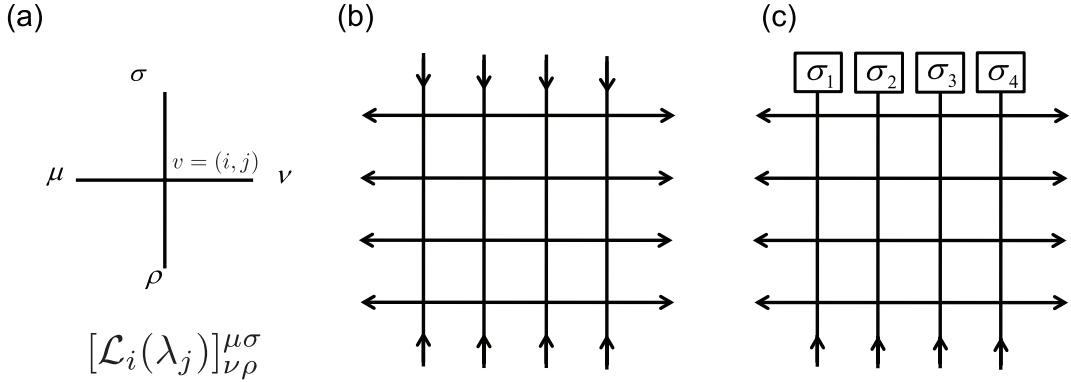}
\vspace{.0cm}
\caption{(a) $L$ operator associated with the vertex which is the intersection of the $i$th vertical line and $j$th horizontal line. (b) Domain wall boundary conditions. (c) Other boundary conditions. Each $\sigma_i$ is $\ua$ or $\da$. }
\label{fig: 6v_DWBC}
\end{center}
\end{figure}

In this section, we shall clarify the relation between the matrix product Bethe ansatz and the six-vertex model with domain wall boundary conditions. 
The six vertex model is a two-dimensional statistical mechanics model in which the Boltzmann weights are assigned to the six different configurations of arrows around a vertex. 
If the Boltzmann weights satisfy the Yang-Baxter relation, the model is exactly solvable by the Bethe ansatz. 
The partition function of this model on a $n \times L$ rectangle is defined by
\bege
{\mathscr Z} = \left[ \prod^n_{j=1} \sinh \left( \lam_j+\frac{\eta}{2} \right) \right]^L Z
~~~~~{\rm with}~~
Z=\sum_{\rm config} \prod_{v=(i,j)} [{\cal L}_i (\lam_j)] ^{\mu_v \sigma_v}_{\nu_v \rho_v},
\ende
where the summation is taken over all the possible configurations satisfying the {\it ice rule} and 
$v=(i,j)$ denotes the vertex which is the intersection of the $i$th vertical and $j$th horizontal lines. 
In the definition of $Z$, the product is taken over all the vertices.  
The Boltzmann weights are related to the six nonzero matrix elements of the $L$-operator and
${\cal L}_i (\lam_j)$ is assigned to the vertex $v$. The indices $\mu$ and $\nu$ correspond to arrows on the horizontal edges while $\sigma$ and $\rho$ correspond to spins on the vertical edges (see Fig. \ref{fig: 6v_DWBC} (a)). 
The model with periodic boundary conditions was first solved by Lieb~\cite{Lieb_6v}. 
The model with domain wall boundary conditions (see Fig. \ref{fig: 6v_DWBC} (b)), which is relevant to our study, was first discovered by Korepin in the context of the norm of the Bethe state~\cite{Korepin_DWBC}. 
For a finite ($n \times n$) lattice, the partition function apart from an overall factor is given by a component of the Bethe state as
\bege
Z^{\rm DWBC}_n = \langle \Downarrow \ket{\lam_1,\ldots,\lam_n},
\ende
where $\langle \Downarrow| = \bigotimes^n_{i=1} {}_i\langle \downarrow |$. 
In Ref.~\cite{Korepin_DWBC}, the recursion relation for $Z^{\rm DWBC}_n$ was derived, and then later it was solved by a determinant formula~\cite{Izergin_DWBC}. Note that in the original papers, an inhomogeneous generalization of the six-vertex model was studied. 
Not only the component of the state $\langle \Downarrow|$ but also the other components of the Bethe state can be interpreted as partition functions of the six-vertex model. The coefficient of the state $|\sigma^{[V_1]}_1, ..., \sigma^{[V_L]}_L \ra$, i.e., 
\bege
\la \sigma^{[V_1]}_1, ..., \sigma^{[V_L]}_L| \lam_1, ..., \lam_n \ra
\ende
corresponds to the partition function with the boundary conditions shown in Fig. \ref{fig: 6v_DWBC} (c) 
(A more detailed discussion is provided in Appendix B. 1). 
Here, $\sigma_i$ ($i=1, 2, ..., L$) denotes the spin state, i.e., $\ua$ or $\da$. 
In this sense, to obtain the Bethe state is equivalent to obtaining the partition function of the six-vertex model with various boundary conditions. 
It appears to be a formidable combinatorial task. However, if we construct a vertex model corresponding to the new basis in ${\bar \mathscr{H}}$, 
this complexity is greatly reduced as shown in Appendix B.4 and 5. 
It corresponds exactly to the similarity transformation from the triangular matrices $D_n$ and $C_n$ to the matrices 
${\cal D}_n$ and ${\cal C}_n^{(i)}$ whose actions are diagonal and almost diagonal except for the single space $\AS_i$, respectively. 
It is important to stress here that this change of basis does not alter $Q_n$, i.e., the domain wall boundary condition in ${\bar \mathscr{H}}$. 
In the corresponding vertex model, the Boltzmann weight of one of six vertices becomes zero. We henceforth call this model a {\it five-vertex model}. 
Note that since the partition function itself is the same after the transformation, the statistical model is unchanged. 
Compared with the six-vertex model, however, the combinatorial complexity has been already resolved in the five-vertex model and hence 
the possible number of configurations involved in the calculation of $Z$ is greatly reduced. 

A similar reduction has been known as $F$-matrices in the context of the Drinfel'd twist and triangular Hopf algebras~\cite{q-alg.9612012, Kitanine-Maillet-Terras} which have been used to calculate the correlation functions in the Heisenberg chain~\cite{Kitanine-Maillet-Terras1, Kitanine-Maillet-Slavnov-Terras, Kitanine-Maillet-Slavnov-Terras1, Kitanine-Maillet-Slavnov-Terras2}. 
The $F$-matrix is named after a factorization of the $R$-matrix and defines a transformation from $\Aa, \Ba, \Ca$, and $\Da$ to the new operators.  By this transformation, $\Aa$ and $\Da$ are transformed into diagonal matrices. On the other hand, $\Ba$ and $\Ca$ are transformed into the matrices whose actions are almost diagonal in $\mathscr{H}$. By expanding the physical spin operators in terms of those diagonal and almost diagonal operators, it is able to handle calculations of various correlation functions. 
Compared with this approach, the matrix $F_n$ we found is a counterpart of the $F$-matrix in $\bar {\mathscr H}$. 
Note that the original $F$-matrix is defined in the physical space $\mathscr H$ while our $F_n$ acts on $\bar {\mathscr H}$. 
It is surprising that the connection between the MPA and the ABA we found turns out to be related to the $F$-matrices. 
It would be interesting to bridge over these two approaches more concretely. 

\section{Conclusion}
In conclusion, we have derived the matrix product representation of the Bethe ansatz state for the XXX and XXZ spin-$\frac{1}{2}$ 
Heisenberg chains from the algebraic Bethe ansatz. 
We have also shown that the finite dimensional representations for the matrices appearing in the matrix product ansatz proposed by Alcaraz and Lazo are equivalent to those obtained from the algebraic Bethe ansatz by use of the nontrivial change of basis in ${\bar {\mathscr H}}$, which is related to the $F$-matrices. 
In the new basis, the matrices have a quite simple structure and the algebraic relations between matrices can be shown very easily. 
The relation between the MPA and the six-vertex model with domain wall boundary conditions has also been discussed. 
It would be of great interest to apply the obtained explicit matrices to calculations of static and dynamical correlation functions~\cite{Korepin-Izergin-Essler-Uglov, Kitanine-Maillet-Terras, Kitanine-Maillet-Slavnov-Terras, Kitanine-Maillet-Slavnov-Terras1, Kitanine-Maillet-Slavnov-Terras2, Sakai_K, Caux-Maillet, Caux-Hagemans-Maillet} and entanglement properties~\cite{Sato-Shiroishi-Takahashi, Weston, Calabrese} in the Heisenberg spin chains. 
The first step should be a calculation of the norm of the Bethe eigenstate, which has been obtained and is related to the determinant expression for the partition function of the six-vertex model with domain wall boundary conditions~\cite{Korepin_DWBC, Izergin_DWBC, Korepin_ZinnJustin}. 
It would also be interesting to find a relation between the matrix product ansatz and the hidden Grassmann structure in the XXZ model discussed in the context of mathematically rigorous approaches~\cite{Boos-Jimbo-Miwa-Smirnov-Takeyama1, Boos-Jimbo-Miwa-Smirnov-Takeyama2, Jimbo-Miwa-Smirnov1}. 
Another interesting direction is a systematic construction of the matrix product state representation of the Bethe states for other integrable models, especially for correlated electron systems such as the Hubbard model~\cite{Lieb-Wu, Essler-Korepin, Takahashi_text, Hubbard_text}. 
Our discussion in Section 2 indicates that it can be obtained if the model is exactly solvable by the algebraic Bethe ansatz. 
However, it is highly nontrivial whether or not we can obtain explicit expressions for the matrices since we need to find a new basis in which the matrices are almost diagonal.

\section*{Acknowledgments}
The authors are grateful to Z. C. Gu, V. E. Korepin, C. Matsui, and F. Verstraete for their valuable comments and discussions. 
After the completion of this work, we learned that a similar study has been done by V. E. Korepin and F. Verstraete~\cite{Korepin-Verstraete}. 
This work was supported in part by Grant-in-Aids (No. 20740214) from the Ministry of Education, Culture, Sports, Science and Technology of Japan. HK is supported by the JSPS Postdoctral Fellow for Research Abroad. 

\appendix
\section{index of notation}
Here is a list of the notations which are used in the main text. 
\begin{itemize}
\item Linear spaces
 \begin{itemize}
 \item $V$: a physical Hilbert space spanned by $\ket{\uparrow}$ and $\ket{\downarrow}$
 \item $\AS$ : an auxiliary space spanned by $\ket{\rightarrow}$ and $\ket{\leftarrow}$
 \item $\mathscr{H}$: $V^{\otimes L}$
 \item ${\bar \mathscr{H}}$ : ${\bar V}^{\otimes n}$
 \end{itemize}
\item{States}
 \begin{itemize}
 \item $\ket{\Uparrow}$: $\ket{\ua, \ua \ldots \ua}$ ($= \Phi_0$)
 \item $\ket{\Downarrow}$: $\ket{\da, \da \ldots \da}$
 \item $\ket{\Leftarrow}$: $\ket{\leftarrow, \leftarrow \ldots \leftarrow}$
 \item $\ket{\Rightarrow}$: $\ket{\rightarrow, \rightarrow \ldots \rightarrow}$
 \end{itemize}
\item Parameters
  \begin{itemize}
\item $b$ (or $z$) $c$ : elements of $L$-operator
\item $\Delta, \eta$: parameters of Hamiltonian
\item $\lambda$: a spectral parameter
\end{itemize}
\item Operators acting on the physical space ($\mathscr{H}$)
  \begin{itemize}
\item $\Aa, \Ba, \Ca, \Da$ : elements of a monodromy matrix used in ABA
\item $\HM$ : Hamiltonian matrix
\item $\TM$ : a transfer matrix
\end{itemize}
\item Operators acting on the auxiliary space ($\bar{\mathscr{H}}$)
  \begin{itemize}
\item $\Cm, \Dm$: matrices used in MPS
\item $\Cf, \Df$: matrices used in the new basis after transformation 
\item $F$ : a similarity transformation
\item ${\cal F}$ : a matrix defined by Eq. (\ref{Ff_n})
\item $R$ : $R$-matrix
\item $Q, {\cal Q}$ : matrices for DWBC. (They turned out to be the same). 
\item $A,E$: matrices used in MPA by Alcaraz and Lazo~\cite{Alcaraz1, Alcaraz2}
\end{itemize}
\item Other operators
  \begin{itemize}
\item ${\cal L}$ : $L$-operator
\item ${\cal T}$ : a monodromy matrix
\end{itemize}
\end{itemize}
\section{Graphical representation\label{sec:graph}}
\subsection{definitions of graphs}
The $L$-operator ${\cal L}_{ji}(\lam_j)$ given by Eq.~(\ref{eq:L-op}) is a $4 \times 4$ matrix acting on a vector space $\AS_j \otimes V_i$. 
The space $\AS_j$ is spanned by $|\leftarrow\ra_j$ and $|\rightarrow \ra_j$ while $V_i$ is by $|\ua\ra_i$ and $|\da\ra_i$. 
We hereafter denote it as ${\cal L}_{i}(\lam_j)$. 
The order of the basis in $\AS_j \otimes V_i$ 
is fixed as $\ket{\leftarrow\uparrow}
,\ket{\leftarrow\downarrow}
,\ket{\rightarrow\uparrow}
,\ket{\rightarrow\downarrow}$. 
In a graphical representation, 
we denote a matrix element as 
\begin{eqnarray*}
\bra{\mu \sigma}{\cal L}_i(\lam_j)\ket{\nu \rho} =[{\cal L}_i (\lam_j)]^{\mu \sigma}_{\nu \rho}
=    \begin{minipage}{0.08\linewidth}
      \includegraphics{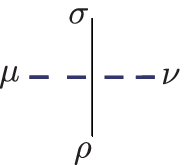}
    \end{minipage}
,
\end{eqnarray*}
where $\mu$ and $\nu$ are $\leftarrow$ or $\rightarrow$ while $\sigma$ and $\rho$ are $\ua$ or $\da$. 
The space $V_i$ (or $\AS_j$) is denoted by a black thick line (a blue bold dashed line). 
There are $2^4=16$ possible configurations of vertices. 
However, 6 of them are nonzero as we have seen in Eq. (\ref{eq:L-op}), which corresponds to the six-vertex model. 
The matrix ${\cal L}_i(\lam_j)$ 
is graphically represented as
\begin{eqnarray}
\fl 
{\cal L}_i(\lam_j)
=
\left(\begin{array}{cccc}
1 & 0 & 0 & 0\\
0 & b(\lambda_j) & c(\lambda_j) & 0\\
0 & c(\lambda_j) & b(\lambda_j) & 0\\
0 & 0 & 0 & 1
\end{array}\right)
=
\left(\begin{array}{cccc}
    \begin{minipage}{0.08\linewidth}
      \includegraphics{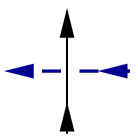}
    \end{minipage}
    & 0 & 0 & 0\\
    0 & 
    \begin{minipage}{0.08\linewidth}
      \includegraphics{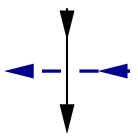}
    \end{minipage}
    & 
    \begin{minipage}{0.08\linewidth}
      \includegraphics{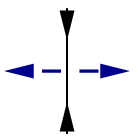}
    \end{minipage}
    & 0\\
    0 & 
    \begin{minipage}{0.08\linewidth}
      \includegraphics{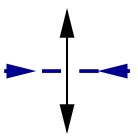}
    \end{minipage}
    & 
    \begin{minipage}{0.08\linewidth}
      \includegraphics{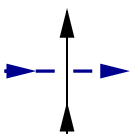}
    \end{minipage}
    & 0\\
    0 & 0 & 0 & 
    \begin{minipage}{0.08\linewidth}
      \includegraphics{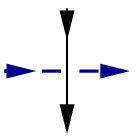}
    \end{minipage}
  \end{array}\right).
\label{geq:L-op}
\end{eqnarray}
The correspondence between the vertices and the ``Boltzmann weights" is summarized as follows: 
\begin{eqnarray*}  
\begin{minipage}{0.08\linewidth}
\includegraphics{xfig/x-lluu.eps}
\end{minipage}
=  
\begin{minipage}{0.08\linewidth}
\includegraphics{xfig/x-rrdd.eps}
\end{minipage}
=1,
\;\;\;  
\begin{minipage}{0.08\linewidth}
\includegraphics{xfig/x-lldd.eps}
\end{minipage}
=  
\begin{minipage}{0.08\linewidth}
\includegraphics{xfig/x-rruu.eps}
\end{minipage}
=b,
\;\;\;
\begin{minipage}{0.08\linewidth}
\includegraphics{xfig/x-lrdu.eps}
\end{minipage}
=  
\begin{minipage}{0.08\linewidth}
\includegraphics{xfig/x-rlud.eps}
\end{minipage}
=c.
\end{eqnarray*}
Note that the weights $b$ and $c$ depend on $j$ and can be complex. 
We should stress here that the above correspondence is not unique but depends on an order of basis.
We now introduce a graphical representation for the $L$-operator. 
The $L$-operator acting on $\AS_j \otimes V_i$ is drawn as a vertex without arrows:
\begin{eqnarray*}
{\cal L}_i(\lam_j)
&=&
\begin{minipage}{0.08\linewidth}
\includegraphics{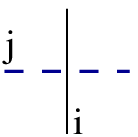}
\end{minipage}
~~.
\end{eqnarray*}
Next we define four kinds of operators acting on $V_i$ from the $L$-operator. 
They are defined by decomposing the $4 \times 4$ matrix in Eq.~(\ref{geq:L-op}) into $2 \times 2$ sub-matrices: 
\begin{eqnarray*}
{\cal L}_i(\lam_j)
=
\begin{minipage}{0.08\linewidth}
\includegraphics{xfig/xij.eps}
\end{minipage}
=
\left(
\begin{array}[]{cc}
\begin{minipage}{0.08\linewidth}
\includegraphics{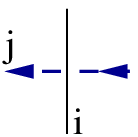}
\end{minipage}
&
\begin{minipage}{0.08\linewidth}
\includegraphics{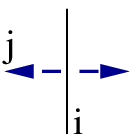}
\end{minipage}
\\
\begin{minipage}{0.08\linewidth}
\includegraphics{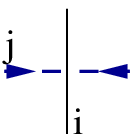}
\end{minipage}
&
\begin{minipage}{0.08\linewidth}
\includegraphics{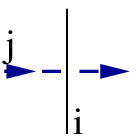}
\end{minipage}
\end{array}
\right)_{\left[\AS_j \right]}
.
\label{geq:L-op:AS}
\end{eqnarray*}
Here, the index $\left[\AS \right]$ denotes the space in which the $L$-operator is represented as a $2 \times 2$ matrix. 
In addition, 
we can define the operators acting on $\AS_j$ 
by a decomposition of the $4\times 4$ matrix: 
\begin{eqnarray*}
\fl 
\left(\begin{array}{cccc}
      1 & 0 & 0 & 0\\
      0 & b(\lambda_j) & c(\lambda_j) & 0\\
      0 & c(\lambda_j) & b(\lambda_j) & 0\\
      0 & 0 & 0 & 1
    \end{array}\right)
  =
  \left(\begin{array}{cccc}
      \begin{minipage}{0.08\linewidth}
        \includegraphics{xfig/x-lluu.eps}
      \end{minipage}
      & 0 & 0 & 0\\
      0 & 
      \begin{minipage}{0.08\linewidth}
        \includegraphics{xfig/x-rruu.eps}
      \end{minipage}
      & 
      \begin{minipage}{0.08\linewidth}
        \includegraphics{xfig/x-rlud.eps}
      \end{minipage}
      & 0\\
      0 & 
      \begin{minipage}{0.08\linewidth}
        \includegraphics{xfig/x-lrdu.eps}
      \end{minipage}
      & 
      \begin{minipage}{0.08\linewidth}
        \includegraphics{xfig/x-lldd.eps}
      \end{minipage}
      & 0\\
      0 & 0 & 0 & 
      \begin{minipage}{0.08\linewidth}
        \includegraphics{xfig/x-rrdd.eps}
      \end{minipage}
    \end{array}\right)
  =
  \left(
    \begin{array}[]{cc}
      \begin{minipage}{0.08\linewidth}
        \includegraphics{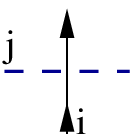}
      \end{minipage}
      &
      \begin{minipage}{0.08\linewidth}
        \includegraphics{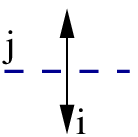}
      \end{minipage}
      \\
      \begin{minipage}{0.08\linewidth}
        \includegraphics{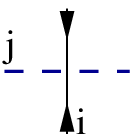}
      \end{minipage}
      &
      \begin{minipage}{0.08\linewidth}
        \includegraphics{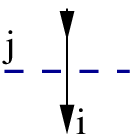}
      \end{minipage}
    \end{array}
  \right)_{\left[V_i\right]}
  .
\end{eqnarray*}
Here, the index $\left[V\right]$ denotes the space in which the $4 \times 4$ matrix is represented as a $2 \times 2$ matrix. 
Although the $4 \times 4$ matrix appearing in the above equation is exactly the same as ${\cal L}_i(\lam_j)$, i.e., the matrix in Eq. (\ref{geq:L-op}), the order of basis has been changed and hence the graphical representations for the matrix elements are different from Eq. (\ref{geq:L-op}). 
The reason why the $4 \times 4$ matrices themselves are the same is related to the fact that 
the Boltzmann weights of the six-vertex model are invariant under the simultaneous reversal of all arrows when there is no external field. 
Two operators out of the four operators acting on $\AS_j$ are defined in Eq.~(\ref{eq:first terms}), 
and graphs of them are given by
\begin{eqnarray*}
D_1(\lambda_j) &=& 
      \begin{minipage}{0.08\linewidth}
        \includegraphics{xfig/xij-00uu.eps}
      \end{minipage}
,\;\;
C_1(\lambda_j) = 
      \begin{minipage}{0.08\linewidth}
        \includegraphics{xfig/xij-00du.eps}
      \end{minipage}
.
\end{eqnarray*}
Note that the absence of arrows on the horizontal (dashed) line indicates that the above operators act on the space $\AS_j$. 

We are now ready to introduce a graphical representation of the Bethe state. 
As was seen in the main text, the Bethe state can be defined through the monodromy matrix 
${\cal T}(\lam_j)$ or ${\cal L}_i(\lam_1,\ldots,\lam_n)$.
These operators are drawn as
\begin{eqnarray}
\fl
  {\cal T}(\lam_j)= \bigotimes_{i=1}^L  {\cal L}_i(\lam_j)
  =
  \prod_{i=1}^L
  \left(
    \begin{array}[]{cc}
      \begin{minipage}{0.08\linewidth}
        \includegraphics{xfig/xij-ll00.eps}
      \end{minipage}
      &
      \begin{minipage}{0.08\linewidth}
        \includegraphics{xfig/xij-lr00.eps}
      \end{minipage}
      \\
      \begin{minipage}{0.08\linewidth}
        \includegraphics{xfig/xij-rl00.eps}
      \end{minipage}
      &
      \begin{minipage}{0.08\linewidth}
        \includegraphics{xfig/xij-rr00.eps}
      \end{minipage}
    \end{array}
  \right)_{\left[\AS_j \right]}
  =
  \begin{minipage}{0.24\linewidth}
    \includegraphics{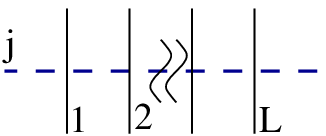}
  \end{minipage}
  ,
  &&
  \\
 \fl 
  {\cal L}_i(\lam_1,\ldots,\lam_n)
  = \bigotimes_{l=0}^{n-1}  {\cal L}_i(\lam_{n-l})
  =
  \prod_{{l=0\atop (j=n-l)}}^{n-1}
  \left(
    \begin{array}[]{cc}
      \begin{minipage}{0.08\linewidth}
        \includegraphics{xfig/xij-00uu.eps}
      \end{minipage}
      &
      \begin{minipage}{0.08\linewidth}
        \includegraphics{xfig/xij-00ud.eps}
      \end{minipage}
      \\
      \begin{minipage}{0.08\linewidth}
        \includegraphics{xfig/xij-00du.eps}
      \end{minipage}
      &
      \begin{minipage}{0.08\linewidth}
        \includegraphics{xfig/xij-00dd.eps}
      \end{minipage}
    \end{array}
  \right)_{\left[V_i\right]}
  =
  \begin{minipage}{0.08\linewidth}
    \includegraphics{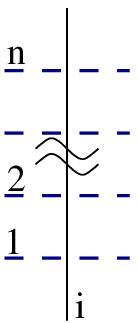}
  \end{minipage}
  \;\;\;.
  &&
  \label{geq:L_n}
\end{eqnarray}
Here, each bond connecting two vertices represents summation over two possible arrows. 
The monodromy matrix is represented as a horinzontal object while ${\cal L}_i(\lam_1, ..., \lam_n)$ is a vertical one. 
The connection between nearest vertices can be easily interpreted as the multiplication of $2 \times 2$ matrices.  
A creation operator for the Bethe state ($\Ba(\lambda_j)$) corresponds to the (1,2)-component of $ {\cal T}(\lambda_j)$ which acts on ${\mathscr H}=V^{\otimes L}$. 
Using $\Ba(\lambda_j)$ represented by 
\begin{eqnarray*}
  \Ba(\lambda_j) &=& \bra{\leftarrow} {\cal T}(\lambda_j) \ket{\rightarrow}
  =
  \begin{minipage}{0.08\linewidth}
    \includegraphics{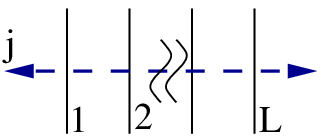}
  \end{minipage}~~~~~~~~~~~~~~~, 
\end{eqnarray*}
the Bethe state is drawn as 
\begin{eqnarray*}
  \ket{\lam_1,\ldots,\lam_n} 
  = \Ba(\lam_n)\cdots \Ba(\lam_1)
  \ket{\Uparrow}
  =
  \begin{minipage}{0.08\linewidth}
    \includegraphics{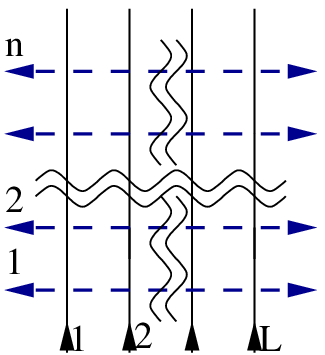}
  \end{minipage}
  \;\;\;\;\;\;\;\;\;\;\;\;\;\;\;\;\;\;.
\end{eqnarray*}
On the other hand, operators $\Dm_n(\lambda_1,\ldots,\lambda_n)=\bra{\uparrow} {\cal L}_i(\lam_1,\ldots,\lam_n) \ket{\uparrow}$
and $\Cm_n(\lambda_1,\ldots,\lambda_n)=\bra{\downarrow} {\cal L}_i(\lam_1,\ldots,\lam_n) \ket{\uparrow}$ on ${\bar {\mathscr H}}=\AS^{\otimes n}$
are represented as
\begin{eqnarray*}
\Dm_n(\lambda_1,\ldots,\lambda_n) &=& 
\begin{minipage}{0.08\linewidth}
\includegraphics{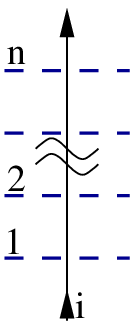}
\end{minipage}
\;,\;\;\;
\Cm_n(\lambda_1,\ldots,\lambda_n) 
=
\begin{minipage}{0.08\linewidth}
\includegraphics{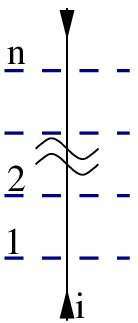}
\end{minipage}
.
\end{eqnarray*}
We shall draw a figure for each component of the Bethe state. 
From the matrix product representation given in Eq. (\ref{eq:BetheStateMPA}), each component is written as
\begin{eqnarray}
\langle \sigma_1^{\left[V_1\right]}
,\ldots,  \sigma_L^{\left[V_L\right]}
\ket{\lam_1,\ldots,\lam_n}
&=
\Tr_{\bar{\mathscr H}} \left[ Q_n \prod_{i=1}^L 
\bra{\sigma_i} {\cal L}_i (\lam_1, ..., \lam_n)\ket{\uparrow}_i
\right]
\nonumber \\ 
&=
\bra{ \Leftarrow}
\prod_{i=1}^L 
\left[
\bra{\sigma_i} {\cal L}_i (\lam_1, ..., \lam_n)\ket{\uparrow}_i
\right]
\ket{\Rightarrow }. 
\label{eq: comp}
\end{eqnarray}
To draw a graph of Eq. (\ref{eq: comp}), we further introduce patches for $\bra{\Leftarrow}$, $\ket{\Rightarrow}$, and $\bra{\sigma_i} {\cal L}_i (\lam_1, ..., \lam_n)\ket{\uparrow}_i$ in the following way: 
\begin{eqnarray*}
\bra{ \Leftarrow }
=
\begin{minipage}{0.08\linewidth}
\includegraphics{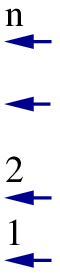}
\end{minipage}
,\;\;\;\;\;  
\bra{\sigma_i} {\cal L}_i (\lam_1, ..., \lam_n)\ket{\uparrow}
=
\begin{minipage}{0.08\linewidth}
\includegraphics{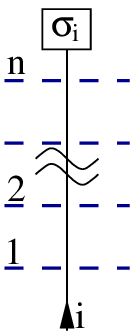}
\end{minipage}
\;\;,\;\;\;\;\;  
\ket{ \Rightarrow}
=
\begin{minipage}{0.08\linewidth}
\includegraphics{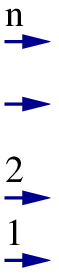}
\end{minipage}
.
&&
\end{eqnarray*}
It is now obvious that the graph for the component of the Bethe state $\ket{\lam_1,\ldots,\lam_n}$ is represented as
\begin{eqnarray*}
\langle \sigma_1^{\left[V_1\right]}
,\ldots,  \sigma_L^{\left[V_L\right]}
\ket{\lam_1,\ldots,\lam_n}
&=&
\begin{minipage}{0.08\linewidth}
\includegraphics{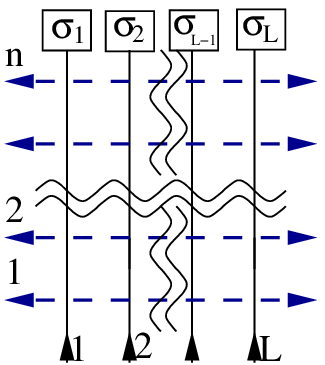}
\end{minipage}~~~~~~~~~~~~~~.   
\end{eqnarray*}
The DWBC in $\mathscr{H}$ corresponds to the case where $\forall \sigma_i=\da$. 
The partition function of the six-vertex model with this boundary condition is graphically given by 
\begin{eqnarray}
Z^{\rm DWBC} 
&=& 
\langle \Downarrow
\ket{\lam_1,\ldots,\lam_n}
=
\begin{minipage}{0.24\linewidth}
\includegraphics{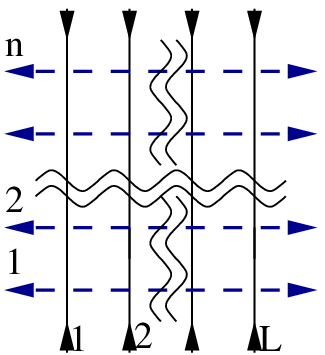}
\end{minipage}
,
\end{eqnarray}
which is the coefficient of the all down (full-filled) state. 
In particular, if $L=n$, it is shown that the partition function is expressed as 
\begin{eqnarray}
Z^{\rm DWBC}_n 
&=
\bra{\Downarrow}
\prod_{l=0}^{n-1} \Ba(\lambda_{n-l})
\ket{\Uparrow}
=
\bra{\Leftarrow}
C_n (\lam_1, ..., \lam_n)^n
\ket{\Rightarrow}
\\
&=
\Tr_{\bar {\mathscr H}}\left[
Q_n 
C_n (\lam_1, ..., \lam_n)^n
\right].
\label{eq:Z_DWBC}
\end{eqnarray}

\subsection{Yang-Baxter equation}
To illustrate the Yang-Baxter equation (YBE) in Eq.~(\ref{eq:RLL=LLR}),
we introduce a graphical representation for the $R$-matrix which is defined by Eq.~(\ref{eq:R-op}):
\begin{eqnarray}
R_{ji}(\lambda_j - \lambda_i) &=& 
=
\begin{minipage}{0.08\linewidth}
\includegraphics{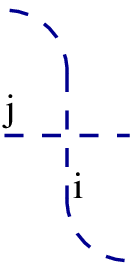}
\end{minipage}
~.
\end{eqnarray}
Here, the parameter $\lambda_i$ (or $\lambda_j$) 
is specified by the space $\AS_i$ ($\AS_j$) for $R_{ji}(\lambda_j - \lambda_i)$. 
Note that this $R$-matrix acts on the space $\AS_j \otimes \AS_i$. 
As shown in Eq.~(\ref{geq:L_n}), ${\cal L}_{i}(\lambda_1) {\cal L}_{i}(\lambda_2)$ is an operator on $V_i \otimes \AS_1 \otimes \AS_2$. Therefore, one can consider a product of this operator with $R_{12}(\lam_1-\lam_2)$. 
The graphical representation for the l.h.s. of the YBE is drawn as
\begin{eqnarray}
R_{12}(\lambda_1 - \lambda_2)
{\cal L}_{i}(\lambda_1) {\cal L}_{i}(\lambda_2)
&=&
\begin{minipage}{0.08\linewidth}
\includegraphics{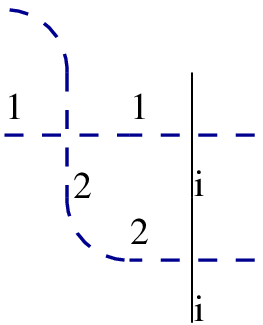}
\end{minipage}
\hspace{0.08\linewidth}, 
\label{geq:YBE1}
\end{eqnarray}
while the r.h.s. is 
\begin{eqnarray}
{\cal L}_{i}(\lambda_2) {\cal L}_{i}(\lambda_1) 
R_{12}(\lambda_1-\lambda_2)
&=&
\begin{minipage}{0.08\linewidth}
\includegraphics{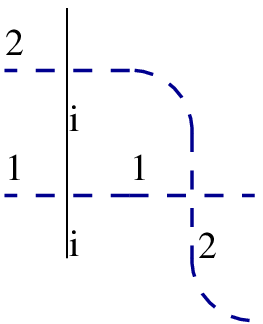}
\end{minipage}
\hspace{0.08\linewidth}.
\label{geq:YBE2}
\end{eqnarray}
Keeping in mind that blue bold dashed lines are attached to the auxiliary spaces, 
we shall henceforth draw pictures for the operators as simply as possible. 
Then the simplified graphical representation of the YBE is given by 
\begin{eqnarray}
\begin{minipage}{0.16\linewidth}
\includegraphics{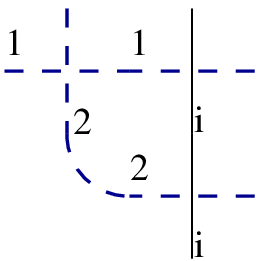}
\end{minipage}
&=&
\begin{minipage}{0.16\linewidth}
\includegraphics{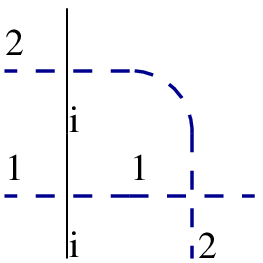}
\end{minipage}
.
\end{eqnarray}
Using this identity repeatedly, one can show the relation Eq.~(\ref{eq:RTT=TTR}) 
which has the following graphical interpretation: 
\begin{eqnarray}
\begin{minipage}{0.32\linewidth}
\includegraphics{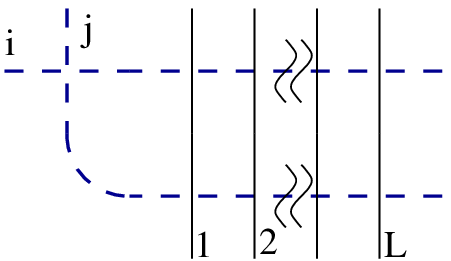}
\end{minipage}
&=&
\begin{minipage}{0.32\linewidth}
\includegraphics{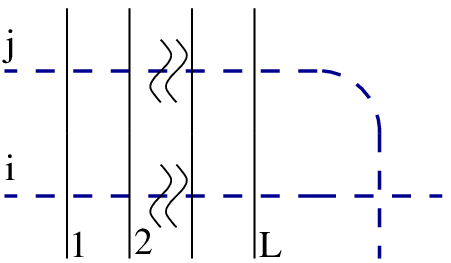}
\end{minipage}
.  
\end{eqnarray}
Note that Eq. (\ref{eq:RTT=TTR}) corresponds to the case where $i=0$ and $j=1$. 

\subsection{Recursion relations in terms of graphs}
In Sec. 3, we have derived the recursion relations for the matrices $D_n$ and $C_n$. 
Those recursions are easy to understand if we use graphical representations.
The recursion relation for $D_n$ (see Eq.~(\ref{recD})),
\begin{equation*}
D_{n+1}=
\left(\begin{array}{cc} 1 & 0 \\ 0 & b(\lam_{n+1}) \end{array} \right)_{[\AS_{n+1}]}
\otimes D_n 
+ \left(\begin{array}{cc} 0 & 0\\ c(\lam_{n+1}) & 0 \end{array}\right)_{[\AS_{n+1}]}
\otimes C_n ,
\end{equation*}
can be expressed as
\begin{equation*}
\hspace{-15mm}
\begin{minipage}{0.08\linewidth}
\includegraphics{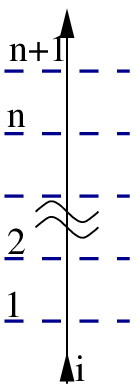}
\end{minipage}
~~~=~~~
\begin{minipage}{0.08\linewidth}
\includegraphics{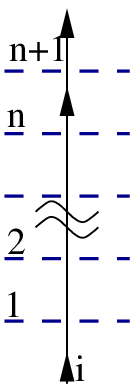}
\end{minipage}
+
\begin{minipage}{0.08\linewidth}
\includegraphics{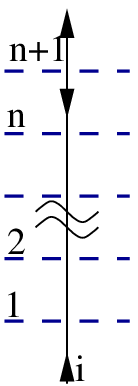}
\end{minipage}~~~
=~~
      \begin{minipage}{0.08\linewidth}
        \includegraphics{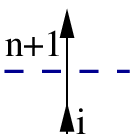}
      \end{minipage}
\times
\begin{minipage}{0.08\linewidth}
\includegraphics{xfig/D.eps}
\end{minipage}
+
      \begin{minipage}{0.08\linewidth}
        \includegraphics{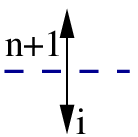}
      \end{minipage}
\times
\begin{minipage}{0.08\linewidth}
\includegraphics{xfig/E.eps}
\end{minipage} ~~. 
\end{equation*}
Here, for simplicity, we have abbreviated the indeterminants $\lam_1, ..., \lam_n$, and $\lam_{n+1}$. 
One can also express the recursion for $C_n$, Eq. (\ref{recE}), in a similar manner: 
\begin{equation*}
\hspace{-15mm}
\begin{minipage}{0.08\linewidth}
\includegraphics{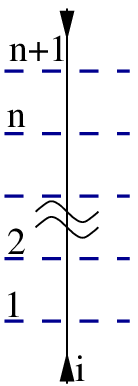}
\end{minipage}
~~~=~~~
\begin{minipage}{0.08\linewidth}
\includegraphics{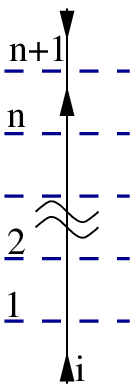}
\end{minipage}
+
\begin{minipage}{0.08\linewidth}
\includegraphics{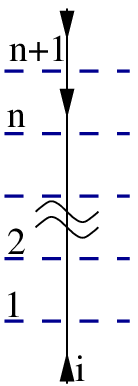}
\end{minipage}
~~~=~~
\begin{minipage}{0.08\linewidth}
\includegraphics{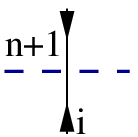}
\end{minipage}
\times
\begin{minipage}{0.08\linewidth}
\includegraphics{xfig/D.eps}
\end{minipage}
+
\begin{minipage}{0.08\linewidth}
\includegraphics{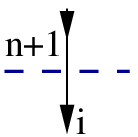}
\end{minipage}
\times
\begin{minipage}{0.08\linewidth}
\includegraphics{xfig/E.eps}
\end{minipage}~~. 
\end{equation*}
The matrix $D_n$ is a lower triangular matrix while $C_n$ is an upper triangular one. 
This can be easily seen by representing them $2 \times 2$ matrices whose elements are graphs: 
\begin{equation}
\fl 
D_{n+1}=
\left(\begin{array}{cc} 
\begin{minipage}{0.08\linewidth}
\includegraphics{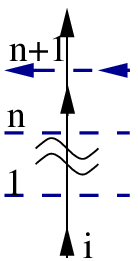}
\end{minipage}
& 
0 \\ 
\begin{minipage}{0.08\linewidth}
\includegraphics{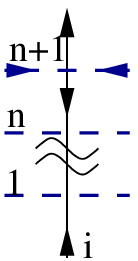}
\end{minipage}
& 
\begin{minipage}{0.08\linewidth}
\includegraphics{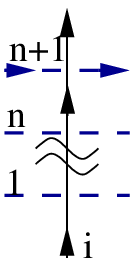}
\end{minipage}
\end{array} \right)_{[\AS_{n+1}]}
,~~~
C_{n+1}=
\left(\begin{array}{cc} 
\begin{minipage}{0.08\linewidth}
\includegraphics{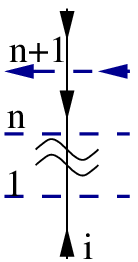}
\end{minipage}
& 
\begin{minipage}{0.08\linewidth}
\includegraphics{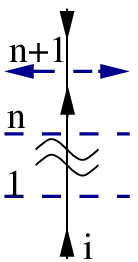}
\end{minipage}
\\ 
0
&
\begin{minipage}{0.08\linewidth}
\includegraphics{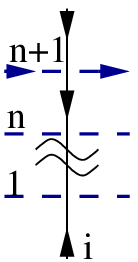}
\end{minipage}
\end{array} \right)_{[\AS_{n+1}]}.
\end{equation}

\subsection{Mapping to five-vertex model}
As we have seen in Sec. 4, the triangular matrices $D_n$ and $C_n$ are transformed into $\Df_n$ and $\Cf_n$ by the invertible matrix $F_n$. This similarity transformation corresponds to a mapping from the six-vertex model to a five-vertex model. 
This relation can be easily understood using graphs. In the new basis, the matrix $\Df_n$ is diagonal and its recursion Eq. (\ref{recD3}) can be represented as
\begin{eqnarray}
\Df_{n+1}&=&
\left(\begin{array}{cc} 
\begin{minipage}{0.08\linewidth}
\includegraphics{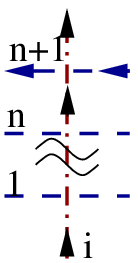}
\end{minipage}
& 
0 \\ 
0
& 
\begin{minipage}{0.08\linewidth}
\includegraphics{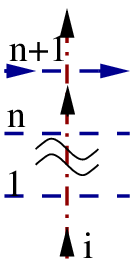}
\end{minipage}
\end{array} \right)_{[\AS_{n+1}]}
~~{\rm with}~~
\Df_{n} =
\begin{minipage}{0.08\linewidth}
\includegraphics{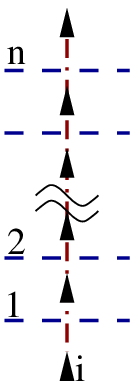}
\label{eq: Df_graph1}
\end{minipage}
,
\end{eqnarray}
where the vertical lines are modified to indicate the transformed operators. 
Since the recursion (\ref{recD3}) uniquely determines the graphical representation for $\Df_n$, all the arrows on the vertical line are $\ua$.  
Note that a horizontal dashed line without arrows acts on the space ($\AS$) diagonally. 
In the main text, the matrix $\Cf_n$ is decomposed as $\Cf_n=\sum^n_{j=1} \Cf^{(j)}_n$. From Eq. (\ref{newCf_2}), we see that $\Cf^{(j)}_n$ flips only the state in $\AS_j$ from $|\rightarrow \rangle$ to $|\leftarrow \rangle$. Therefore, $\Cf^{(j)}_n$ can be written as 
\begin{equation}
\fl ~~~~~~~~
\Cf_n^{(j)}  = \ket{\leftarrow}_j \bra{\leftarrow}\Cf_n^{(j)}\ket{\rightarrow}_j \bra{\rightarrow}
~~~~{\rm with}~~~~
_j\bra{\leftarrow}\Cf_n^{(j)}\ket{\rightarrow}_j
=~~~ 
\begin{minipage}{0.08\linewidth}
\includegraphics{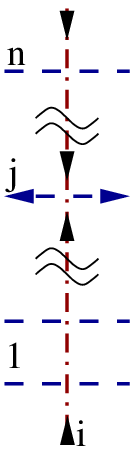}
\end{minipage}.
\end{equation}
Since the action of $\Cf^{(j)}_n$ is diagonal in ${\bar {\mathscr H}}$ except for $\AS_j$, the arrows on the vertical line are specified as follows:
\begin{eqnarray}
_j\bra{\leftarrow}\Cf_n^{(j)}\ket{\rightarrow}_j  =
\begin{minipage}{0.08\linewidth}
\includegraphics{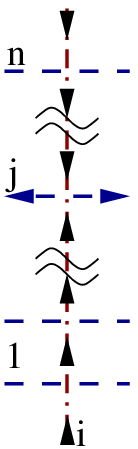}
\end{minipage}.
\label{eq: Cf_graph2}
\end{eqnarray}
From the graphical representation, $\Cf^{(j)}_n$ can be regarded as a kink in ${\bar {\mathscr H}}$. 
It is consistent with the picture that $A_{k_j,n}$ becomes a Jordan-Wigner fermion in ${\bar {\mathscr H}}$ when $\Delta=0$ as we discussed in Sec. 4. 
Again we note that the the horizontal line without arrows acts diagonally on the space. In Eqs. (\ref{eq: Df_graph1}) and (\ref{eq: Cf_graph2}), the vertex
$ \begin{minipage}{0.08\linewidth}\includegraphics{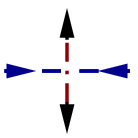}\end{minipage}$ 
does not appear, which means that the original six-vertex model becomes a five-vertex model after the similarity transformation. 
Comparing the graphs with Eqs. (\ref{newDf}) and (\ref{newCf_2}), we find the weights for 6 vertices:
\begin{eqnarray*}
\begin{minipage}{0.08\linewidth}
\includegraphics{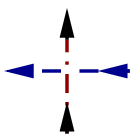}
\end{minipage}
=1  
\;\;\;
,
\begin{minipage}{0.08\linewidth}
\includegraphics{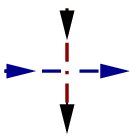}
\end{minipage}
=z_j
,
\;\;\;
\begin{minipage}{0.08\linewidth}
\includegraphics{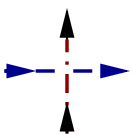}
\end{minipage}
=z_j
,
\;\;\;
\begin{minipage}{0.08\linewidth}
\includegraphics{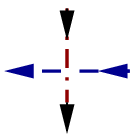}
\end{minipage}
=S_{k,j}(z_k,z_j)
,\\
\begin{minipage}{0.08\linewidth}
\includegraphics{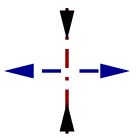}
\end{minipage}
=f(z_j,\ldots,z_n)c(\lambda_j)
,
\;\;\;
\begin{minipage}{0.08\linewidth}
\includegraphics{xfig/x-rlud2.eps}
\end{minipage}
=0,
\end{eqnarray*}
where $j$ and $k$ denote the horizontal line and the index for $\Cf_n^{(k)}$, respectively. 
Note that the weight of the 4th vertex is determined in a non-local way, i.e., 
if the intersection of the vertical line and $k$th horizontal line is 
$\begin{minipage}{0.08\linewidth} \includegraphics{xfig/x-lrdu2.eps} \end{minipage}$, 
then  the weight for the intersection of the vertical line and the $j$th horizontal line with $j > k$ is given by $S_{k,j}(z_k,z_j)$
when both the arrows on the horizontal edges are $\leftarrow$. 
The weight for the 5th vertex also depends nonlocally on $j+1, j+2, ..., n$. 
In exchange for this non-locality, the combinatorial complexity is greatly reduced in the five-vertex model. 
In the original matrices $D_n$ and $C_n$, there are $2^{n-1}$ possible configurations of arrows ($\ua$ and $\da$) on the vertical line. 
After the similarity transformation, however, the configuration for $\Df_n$ is uniquely determined as has seen in Eq. (\ref{eq: Df_graph1}).  
The number of possible configurations of arrows for $\Cf_n$ is $n$ since each $\Cf^{(j)}_n$ is uniquely determined (see Eq. (\ref{eq: Cf_graph2})).  

Finally, we represent $Z^{\rm DWBC}_n$ in terms of the transformed graphs.
Recalling that the DWBC is not altered after the similarity transformation, i.e., 
$Q_n = \ket{ \Rightarrow } \bra{ \Leftarrow } = \Qf_n$, and $F_n F_n^{-1} = 1$,
Eq.~(\ref{eq:Z_DWBC}) is written as
\begin{eqnarray}
Z^{\rm DWBC}_n &=
\Tr_{\bar {\mathscr H}}\left[Q_n C_n (\lam_1, ..., \lam_n)^n \right]
\nonumber \\   
&= \Tr_{\bar {\mathscr H}}\left[
F_n^{-1} Q_n F_n
\left( F_n^{-1} C_n (\lam_1, ..., \lam_n) F_n \right)^n
\right]
\nonumber \\
&=
\Tr_{\bar {\mathscr H}}\left[
{\cal Q}_n {\cal C}_n (\lam_1, ..., \lam_n)^n \right]
\nonumber \\
&=
\bra{\Leftarrow}
{\cal C}_n (\lam_1, ..., \lam_n)^n
\ket{\Rightarrow}. 
\end{eqnarray}
In short, what we have shown is graphically represented as
\begin{eqnarray}
Z^{\rm DWBC}_n&=&
\begin{minipage}{0.24\linewidth}
\includegraphics{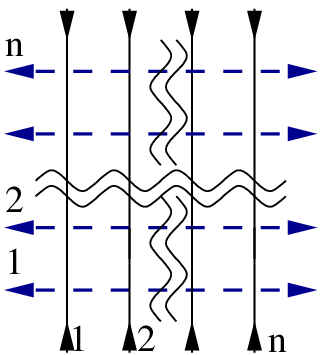}
\end{minipage}
=
\begin{minipage}{0.24\linewidth}
\includegraphics{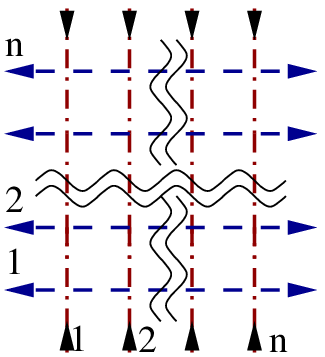}
\end{minipage}
\end{eqnarray}
with 
$\begin{minipage}{0.08\linewidth}
\includegraphics{xfig/x-rlud2.eps}
\end{minipage}=0
$.
Again it should be stressed here that a merit of the five-vertex model introduced is the reduction of possible configurations 
involved in the calculation of $Z^{\rm DWBC}_n$.

\section*{References}
\bibliographystyle{unsrt}
\bibliography{mybib.bib}

\providecommand{\newblock}{}
\begin{thebibliography}{10}
\expandafter\ifx\csname url\endcsname\relax
  \def\url#1{{\tt #1}}\fi
\expandafter\ifx\csname urlprefix\endcsname\relax\def\urlprefix{URL }\fi
\providecommand{\eprint}[2][]{\url{#2}}

\bibitem{Verstraete-Cirac}
Verstraete F and Cirac J~I 2006 {\em Phys. Rev. B\/} {\bf
  {\textbf{73}}} 094423

\bibitem{Perez-Garcia}
Perez-Garcia D, Verstraete F, Wolf M~M and Cirac J~I 2007 {\em Quantum Inf.
  Comput.\/} {\bf {\textbf{7}}} 401

\bibitem{AKLT_PRL}
Affleck I, Kennedy T, Lieb E~H and Tasaki H 1987 {\em Phys. Rev. Lett.\/} {\bf
  \textbf{59}} 799

\bibitem{AKLT_CMP}
Affleck I, Kennedy T, Lieb E~H and Tasaki H 1988 {\em Commun. Math. Phys.\/}
  {\bf \textbf{115}} 477

\bibitem{Baxter}
{A similar concept has also been found by Baxter for classical models in
  statistical mechanics} and Baxter R~J 1978 {\em J. Stat. Phys.\/} {\bf
  \textbf{19}} 461

\bibitem{Haldane_PLA}
Haldane F~D~M 1983 {\em Phys. Lett.\/} {\bf \textbf{93A}} 464

\bibitem{Haldane_PRL}
Haldane F~D~M 1983 {\em Phys. Rev. Lett.\/} {\bf \textbf{50}} 1153

\bibitem{Fannes}
Fannes M, Nachtergaele B and Werner R~F 1992 {\em Commun. Math. Phys.\/} {\bf
  {\textbf{144}}} 443

\bibitem{Kluemper1}
Kl\"umper A, Schadschneider A and Zittarz J 1992 {\em Z. Phys. B\/} {\bf
  {\textbf{87}}} 281

\bibitem{Kluemper2}
Kl\"umper A, Schadschneider A and Zittarz J 1993 {\em Europhys. Lett.\/} {\bf
  {\textbf{24}}} 293

\bibitem{White1}
White S~R 1992 {\em Phys. Rev. Lett.\/} {\bf {\textbf{69}}} 2863

\bibitem{White2}
White S~R 1993 {\em Phys. Rev. B\/} {\bf {\textbf{48}}} 10345

\bibitem{Rommer}
\"Ostlund S and Rommer S 1995 {\em Phys. Rev. Lett.\/} {\bf {\textbf{75}}} 3537

\bibitem{Vidal}
Vidal G, Lattore J~I, Rico E and Kitaev A 2003 {\em Phys. Rev. Lett.\/} {\bf
  {\textbf{90}}} 227902

\bibitem{Alcaraz1}
Alcaraz F~C and Lazo M~J 2004 {\em J. Phys. A: Math. Gen.\/} {\bf
  {\textbf{37}}} L1

\bibitem{Alcaraz2}
Alcaraz F~C and Lazo M~J 2006 {\em J. Phys. A: Math. Gen.\/} {\bf
  {\textbf{39}}} 11335

\bibitem{Bethe}
Bethe H~A 1931 {\em Z. Phys.\/} {\bf {\textbf{71}}} 205

\bibitem{Zam}
Zamolodchikov A~B and Zamolodchikov A~B 1979 {\em Annals of Phys., NY\/} {\bf
  {\textbf{120}}} 253

\bibitem{Alcaraz-Lazo3}
Alcaraz F~C and Lazo M~J 2004 {\em J. Phys. A: Math. Gen.\/} {\bf
  {\textbf{37}}} 4149

\bibitem{Lieb-Wu}
Lieb E~H and Wu F~Y 1968 {\em Phys. Rev. Lett.\/} {\bf \textbf{20}} 1445

\bibitem{Schlottmann}
Schlottmann P 1987 {\em Phys. Rev. B\/} {\bf \textbf{36}} 5177

\bibitem{Fateev-Zamolodchikov}
Zamolodchikov A~B and Fateev V 1980 {\em Sov. J. Nucl. Phys.\/} {\bf
  \textbf{32}} 298

\bibitem{Faddeev}
Faddeev L~D 1984 {\em Les Houches 1982, Recent Advances in Field Theory and
  Statistical Mechanics\/} (Elsevier Science, Amsterdam) p 561

\bibitem{QISM}
Korepin V~E, Bogoliubov N~M and Izergin A~G 1993 {\em Quantum Inverse
  Scattering Method and Correlation Functions \/}
  (Cambridge University Press)

\bibitem{Nepomechie}
Nepomechie R~I 1999 {\em Int. J. Mod. Phys. B\/} {\bf {\textbf{13}}} 2973
  (hep--th/9810032)

\bibitem{Golinelli-Mallick}
Golinelli O and Mallick K 2006 {\em J. Phys. A: Math. Gen.\/} {\bf
  {\textbf{39}}} 10647

\bibitem{Derrida}
Derrida B, Evans M~R, Hakim V and Pasquier V 1993 {\em J. Phys. A: Math.
  Gen.\/} {\bf {\textbf{26}}} 1493

\bibitem{AlcarazDroz}
Alcaraz F~C, Droz M, Henkel M and Rittenberg V 1994 {\em Ann. Phys., NY\/} {\bf
  230} 250

\bibitem{Korepin_DWBC}
Korepin V~E 1982 {\em Commun. Math. Phys.\/} {\bf {\textbf{86}}} 391

\bibitem{Deguchi}
Deguchi T 2003 {\em Classical and Quantum Nonlinear Integrable Systems: Theory
  and Applications\/} (Institute of Physics Publishing) p 113

\bibitem{Lieb_6v}
Lieb E~H 1967 {\em Phys. Rev.\/} {\bf \textbf{162}} 162

\bibitem{Izergin_DWBC}
Izergin A~G 1987 {\em Sov. Phys. Doklady\/} {\bf {\textbf{32}}} 878

\bibitem{q-alg.9612012}
Maillet J~M and de~Santos J~S 1996 {\em q-alg/9612012\/} {\bf {\textbf{}}}

\bibitem{Kitanine-Maillet-Terras}
Kitanine N, Maillet J~M and Terras V 1999 {\em Nucl. Phys. B\/} {\bf
  {\textbf{554}}} 647

\bibitem{Kitanine-Maillet-Terras1}
Kitanine N, Maillet J~M and Terras V 2000 {\em Nucl. Phys. B\/} {\bf
  {\textbf{567}}} 554

\bibitem{Kitanine-Maillet-Slavnov-Terras}
Kitanine N, Maillet J~M, Slavnov N~A and Terras V 2002 {\em Nucl. Phys. B\/}
  {\bf {\textbf{641}}} 487

\bibitem{Kitanine-Maillet-Slavnov-Terras1}
Kitanine N, Maillet J~M, Slavnov N~A and Terras V 2005 {\em Nucl. Phys. B\/}
  {\bf {\textbf{712}}} 600

\bibitem{Kitanine-Maillet-Slavnov-Terras2}
Kitanine N, Maillet J~M, Slavnov N~A and Terras V 2005 {\em Nucl. Phys. B\/}
  {\bf {\textbf{729}}} 558

\bibitem{Korepin-Izergin-Essler-Uglov}
Korepin V~E, Izergin A~G, Essler F~H~L and Uglov D~B 1994 {\em Phys. Lett\/}
  {\bf {\textbf{A190}}} 182

\bibitem{Sakai_K}
Sakai K 2007 {\em J. Phys. A: Math. Theor.\/} {\bf {\textbf{40}}} 7523

\bibitem{Caux-Maillet}
Caux J~S and Maillet J~M 2005 {\em Phys. Rev. Lett.\/} {\bf {\textbf{95}}}
  077201

\bibitem{Caux-Hagemans-Maillet}
Caux J~S, Hagemans R and Maillet J~M 2005 {\em J. Stat. Mech\/} {\bf
  {\textbf{}}} P09003

\bibitem{Sato-Shiroishi-Takahashi}
Sato J, Shiroishi M and Takahashi M 2006 {\em J. Stat. Mech.\/} {\bf
  {\textbf{}}} P12017

\bibitem{Weston}
Weston R 2006 {\em J. Stat. Mech.\/} {\bf {\textbf{}}} L03002

\bibitem{Calabrese}
Alba V, Fagotti M and Calabrese P 2009 {\em J. Stat. Mech.\/} {\bf {\textbf{}}}
  P10020

\bibitem{Korepin_ZinnJustin}
Korepin V~E and Zinn-Justin P 2000 {\em J. Phys. A: Math. Gen.\/} {\bf
  {\textbf{33}}} 7053

\bibitem{Boos-Jimbo-Miwa-Smirnov-Takeyama1}
Boos H, Jimbo M, Miwa T, Smirnov F and Takeyama Y 2007 {\em Commun. Math.
  Phys.\/} {\bf {\textbf{272}}} 263

\bibitem{Boos-Jimbo-Miwa-Smirnov-Takeyama2}
Boos H, Jimbo M, Miwa T, Smirnov F and Takeyama Y 2009 {\em Commun. Math.
  Phys.\/} {\bf {\textbf{286}}} 875

\bibitem{Jimbo-Miwa-Smirnov1}
Jimbo M, Miwa T and Smirnov F 2009 {\em J. Phys. A: Math. Theor.\/} {\bf
  {\textbf{42}}} 304018

\bibitem{Essler-Korepin}
Korepin V~E and Essler F~H~L 1994 {\em Exactly Solvable Models of Strongly
  Correlated Electrons\/} (World Scientific, Singapore)

\bibitem{Takahashi_text}
Takahashi M 1999 {\em Thermodynamics of One-Dimensional Solvable Models\/}
  (Cambridge University Press)

\bibitem{Hubbard_text}
Essler F~H~L, Frahm H, G\"ohmann F, Kl\"umper A and Korepin V~E 2005 {\em The
  One-Dimensional Hubbard Model\/} (Cambridge University Press)

\end{thebibliography}

\end{document}